\input harvmac
\writedefs
\sequentialequations

\def \hal {{1\ov 2}}

\def\tf#1#2{{\textstyle{#1 \over #2}}}   
\def\df#1#2{{\displaystyle{#1 \over #2}}}

\def\+{^\dagger}
\def\d{d}
\def\e{e}
\def\i{i}

\def\TL{\hfil$\displaystyle{##}$}
\def\TR{$\displaystyle{{}##}$\hfil}

\def\TT{\hbox{##}}

\def \om {\omega}

\def \do {\dot}
\def\H {{\cal H}}

\def \Q {{\hat Q}}
\def \P {{\hat P}}
\def \q {{\hat q}}

\def \k {\kappa} 

\def \g {\gamma}
\def \del {\partial}

\def \const {{\rm const}}

\def \a {\alpha}
\def \b {\beta}
\def \chi {\chi}\def\r {\rho}
\def \s {\sigma}
\def \p {\phi}
\def \m {\mu}
\def \n {\nu}

\def \l {\lambda}
\def \t {\tau}
\def \td {\tilde }

\def \o {\omega}
\def \inv {^{-1}}
\def \ov {\over }

\def \fourth{{{1\over 4}}}
\def \ha {{1\ov 2}}
\def \QQ {{\cal Q}}

\def \lr { \lref}
\def\np {{  Nucl. Phys. }}
\def \pl {{  Phys. Lett. }}
\def \mpl {{ Mod. Phys. Lett. }}
\def \prl {{  Phys. Rev. Lett. }}
\def \pr  {{ Phys. Rev. }}


\Title{
 \vbox{\baselineskip10pt
  \hbox{PUPT-1650}
  \hbox{CERN-TH/96-270}
  \hbox{Imperial/TP/95-96/71}
  \hbox{hep-th/9610172}
 }
}
{
 \vbox{
  \centerline{Absorption of Fixed Scalars and}
  \vskip 0.1 truein
  \centerline{the D-brane Approach to Black Holes }
 }
}
\vskip -25 true pt

\centerline{
 C.G.~Callan,~Jr.,\footnote{$^1$}{e-mail:  callan@viper.princeton.edu}
 S.S.~Gubser,\footnote{$^2$}{e-mail:  ssgubser@puhep1.princeton.edu}
 I.R.~Klebanov,\footnote{$^3$}{e-mail:  klebanov@puhep1.princeton.edu} }
\centerline{\it Joseph Henry Laboratories, 
Princeton University, Princeton, NJ  08544}

\centerline{ and}

\centerline{
 A.A.~Tseytlin\footnote{$^4$}{e-mail:  tseytlin@ic.ac.uk}\footnote
  {$^{\dagger}$}{On leave from Lebedev Physics Institute, Moscow.} }
\centerline{{\it Theory Division, CERN, Geneve\   and  \  Blackett 
             Laboratory,  Imperial College,  London}}

\centerline {\bf Abstract}
\smallskip
\baselineskip10pt
\noindent

We calculate the emission and absorption rates of fixed scalars by the
near-extremal five-dimensional black holes that have recently been
modeled using intersecting D-branes. We find agreement between the
semi-classical and D-brane computations. At low energies the fixed
scalar absorption cross-section is smaller than for ordinary scalars
and depends on other properties of the black hole than just the
horizon area. In the D-brane description, fixed scalar absorption is
suppressed because these scalars must split into at least four, rather
than two, open strings running along the D-brane. Consequently, this
comparison provides a more sensitive test of the effective string
picture of the D-brane bound state than does the cross-section for
ordinary scalars. In particular, it allows us to read off the value of
the effective string tension. That value is precisely what is needed
to reproduce the near-extremal 5-brane entropy.

\Date {October 1996}

\noblackbox
\baselineskip 14pt plus 1pt minus 1pt


\lref\mast{J.M.~Maldacena and A.~Strominger, Rutgers preprint RU-96-78, 
hep-th/9609026.}
\lref\mst{J.M.~Maldacena and A.~Strominger, \prl 77 (1996) 428, 
hep-th/960.}

\lref\gkgrey{S.S.~Gubser and I.R.~Klebanov, Princeton preprint
PUPT-1648, hep-th/9609076.}

\lref\kr{B.~Kol and A.~Rajaraman, Stanford preprint SU-ITP-96-38, 
SLAC-PUB-7262, hep-th/9608126.}

\lref\gkk{G.~Gibbons, R.~Kallosh and B.~Kol, Stanford preprint 
SU-ITP-96-35, hep-th/9607108.}

\lref\fkk{S. Ferrara and  R. Kallosh, hep-th/9602136; hep-th/9603090;
S. Ferrara, R. Kallosh, A. Strominger, Phys. Rev. D{52} (1995)
5412, hep-th/9508072. }

\lref\hmf{{\it Handbook of Mathematical Functions}, 
M.~Abramowitz and I.A.~Stegun, eds. (US Government Printing Office,
Washington, DC, 1964) 538ff.}

\lref\GR{I.S.~Gradshteyn and I.M.~Ryzhik, {\it Table of Integrals, 
Series, and Products}, Fifth Edition, A.~Jeffrey, ed. (Academic Press:
San Diego, 1994).}

\lref\Unruh{W.G.~Unruh, Phys.~Rev.~{D}14 (1976) 3251.}

\lref\sv{A.~Strominger and C.~Vafa, \pl B379 (1996) 99, hep-th/9601029.}

\lref\cm{C.G.~Callan and J.M.~Maldacena, \np B472 (1996) 591, hep-th/9602043.}

\lref\ms{J.M.~Maldacena and L.~Susskind, Stanford preprint
SU-ITP-96-12, hep-th/9604042.}

\lref\dmw{A.~Dhar, G.~Mandal and S.~R.~Wadia, 
Tata preprint 
TIFR-TH-96/26, hep-th/9605234.}

\lref\dm{S.R.~Das and S.D.~Mathur, 
 hep-th/9606185; hep-th/9607049.}

\lref\dmI{S.R.~Das and S.D.~Mathur,  hep-th/9601152.}

\lref\hms{G.~Horowitz, J.~Maldacena and A.~Strominger, \pl B383 (1996) 151,
hep-th/9603109.}

\lref\HM{G.~Horowitz  and A.~Strominger, \prl  77 (1996) 2368, 
hep-th/9602051.}

\lref\gkt{J.P.~Gauntlett, D.~Kastor and J.~Traschen,
hep-th/9604179.}

\lref\us{S.S.~Gubser and I.R.~Klebanov, hep-th/9608108.}

\lref\dgm{S.~Das, G.~Gibbons and S.~Mathur, hep-th/9609052.}

\lref\ktI{I.R.~Klebanov and A.A.~Tseytlin, Princeton  
preprint PUPT-1639, hep-th/9607107.}

\lref\KT{I.R. Klebanov and A.A. Tseytlin, \np B475 (1996) 179, 
hep-th/9604166.}

\lref\at{ A.A. Tseytlin, \np B475 (1996) 149, hep-th/9604035.}

\lref\ATT{A.A. Tseytlin, \mpl A11 (1996) 689,  hep-th/9601177.}

\lref\CY{M. Cveti\v c and D. Youm, \pr D53 (1996) 584, hep-th/9507090;
Contribution to `Strings 95',  hep-th/9508058. }

\lref\CT{M. Cveti\v c and  A.A.  Tseytlin, \np B478 (1996) 181,
hep-th/9606033.}

\lref\CTT{M. Cveti\v c and  A.A.  Tseytlin, \pr D53 (1996) 5619, 
 hep-th/9512031.}

\lref\CYY{M. Cveti\v c and D. Youm, hep-th/9603100.} 

\lref\juan{J. Maldacena, hep-th/9605016.}

\lref\kaaa{ R. Kallosh, A. Linde, T. Ort\'in, A. Peet and A.  Van
Proeyen, Phys. Rev. D{46} (1992) 5278.}

\lref\myers{C. Johnson, R. Khuri and R. Myers, hep-th/9603061,
Phys. Lett. B378 (1996) 78.} 

\lref\gibb{G. Gibbons, Nucl. Phys. {B207} (1982)  337;
P. Breitenlohner, D. Maison and G. Gibbons, Commun. Math. Phys.
{ 120} (1988) 295.}

\lref\HS{G.T. Horowitz  and A. Strominger, \prl 77 (1996) 2368,
hep-th/9602051.}

\lref\LWW {F. Larsen and F. Wilczek,  \pl B375 (1996) 37,
hep-th/9511064; hep-th/9609084.}

\lref\age{D. Page, \pr D13 (1976) 198.}

\lref\maha{J. Maharana and J.H. Schwarz, \np B390 (1993) 3,
hep-th/9207016.}

\lr \brek{J.C. Breckenridge, R.C. Myers, A.W. Peet and C. Vafa,
hep-th/9602065;
J.C. Breckenridge, D.A. Lowe, R.C. Myers, 
 A.W. Peet, A. Strominger and C. Vafa,
\pl B381 (1996) 423, hep-th/9603078.}

\lr \pertuu{
G. Gilbert,  hep-th/9108012;
 C.F.E. Holzhey  and F. Wilczek, Nucl. Phys. B380 (1992) 447, 
 hep-th/9202014; R. Gregory and  R. Laflamme,  Phys. Rev. D51 (1995) 305,
 hep-th/9410050;
 Nucl. Phys. B428 (1994) 399, 
 hep-th/9404071.}

\lr \khuri{R. Khuri, \np B376 (1992) 350.}

\lr \schw{J.H. Schwarz, \np B226 (1983) 269.    }
\lr\bho{E. Bergshoeff, C. Hull and T. Ort\'in, \np B451 (1995) 547, hep-th/9504081.} 

\lref\dmI{S.R.~Das and S.D.~Mathur, Phys. Lett. B375 (1996) 103, 
 hep-th/9601152.}

\lref\hrs{E.~Halyo, B.~Kol, A.~Rajaraman and L.~Susskind, hep-th/9609075;
E.~Halyo, hep-th/9610068.}

\lref\tssm{A.A. Tseytlin, Nucl. Phys. B477 (1996) 431, hep-th/9605091.}

\lref\kk{I.R.~Klebanov and M.~Krasnitz, Princeton preprint PUPT-1671,
hep-th/9612051.}

\lr \tsetl {A.A. Tseytlin, hep-th/9609212.}

\lr \kabat {
M.R. Douglas, D. Kabat, P. Pouliot and  S.H. Shenker,
  hep-th/9608024.}

\newsec{Introduction}

Many conventional wisdoms of general relativity
are being reconsidered in the context of string theory simply
because the string effective actions for gravity coupled to matter
are more general than those considered  in the past.  
One of the important differences is the presence  of  non-minimal  
scalar--gauge field couplings, leading to a breakdown of the
`no hair' theorem (see the discussion in  \LWW).
Another new effect is the existence of certain scalars which,
in the presence of an extremal charged black hole with regular 
horizon \refs{\kaaa,\CY,\CTT,\CYY}, 
acquire an effective
potential \refs{\gibb} which fixes their value 
at the horizon \refs{\fkk,\gkk}. These 
are the fixed scalars. The absorption of fixed scalars into 
$D=4$ extremal black holes was recently 
considered in \kr\ and found to be suppressed compared to ordinary scalars:
whereas the absorption cross-section of the latter approaches the horizon 
area $A_{\rm h}$ as $\omega\rightarrow 0$ \dgm, the fixed scalar 
cross-section was found to vanish as $\omega^2$. 

The main result of this paper is the demonstration that 
the fixed scalar emission and absorption rates, 
as calculated using the methods
of semi-classical gravity, are exactly reproduced
by the effective string model of black holes based on
intersecting D-branes.  The D-brane description of the five-dimensional 
black holes involves $n_1$ 1-branes and $n_5$ 
5-branes with some left-moving 
momentum along the intersection \refs{\sv,\cm}. 
The low-energy dynamics of the
resulting bound state is believed to be well described by an
effective string wound $n_1 n_5$ times around the compactification
volume \refs{\ms,\dmw,\dm,\us,\mast}. 
This model has been successful in matching not only the
extremal \refs{\sv,\cm} and near-extremal \refs{\HS,\hms,\ms} entropies, but
the rate of Hawking radiation of ordinary
scalars as well \refs{\dm,\us,\mast}.

As part of our study, we have computed the semi-classical absorption
cross-section of fixed scalars from both extremal and
near-extremal $D=5$ black holes. In general, we find
cross-sections with a non-trivial energy dependence. In particular,
for the extremal $D=5$ black holes
with two charges equal, 
$$
\sigma_{\rm abs} = {\pi^2\over 2} R^2 r_K^3 \omega^2
    {{\omega \over 2 T_L} \over 1 - \e^{-{\omega \over 2 T_L}}}
\left (1 + {\omega^2 \over 16 \pi^2 T_L^2} \right )
$$
where $r_K$, $R \gg r_K$ and
$T_L$ are parameters related to the charges.
At low energies the cross-section vanishes as
$\omega^2$, just as in the $D=4$ case studied in
\kr. For non-extremal black holes,
however, the cross-section no longer vanishes as $\omega\rightarrow 0$.
For near-extremal $D=5$ black holes, we find (for
$\omega\sim T_H \ll T_L$)
$$ \sigma_{\rm abs} (\omega)=  {1\over 4} A_{\rm h} r_K^2
(\omega^2+ 4 \pi^2 T_H^2)
\ ,$$
where $T_H$ is the Hawking temperature.
A similar formula holds for the $D=4$ case. Thus,
even at low energies, the fixed
scalar cross-section is sensitive to 
several features of the black hole geometry.
By comparison, the limiting value of the ordinary scalar
cross-section is given by the horizon area alone.
All of the complexities of the fixed scalar emission and absorption
will be reproduced by, and find a simple explanation in,
the effective string picture.

The absorption cross-section for ordinary scalars finds its
explanation in the D-brane description in terms of the process
$scalar \to L + R$ together with its time-reversal $L + R \to scalar$, 
where $L$ and $R$ represent left-moving and right-moving modes
on the effective string \refs{\cm,\dmw,\dm,\us,\mast}. 
The absorption cross-section for fixed scalars is so
interesting because, as we will show, it 
depends on the existence of eight
kinematically permitted processes: 
  \eqn\FixedDProc{\eqalign{
   &1) \ \ scalar \to L + L + R + R \cr
   &2) \ \ scalar + L \to L + R + R \cr
   &3) \ \ scalar + R \to L + L + R \cr
   &4) \ \ scalar + L + R \to L + R \cr
  }}
 and their time-reversals.  One of the main results of this paper is
that competition among $1$--$4$ and their time-reversals gives the
following expression for the fixed scalar
absorption cross-section,
  \eqn\SigmaDbrane{
   \sigma_{\rm abs}(\omega) = { \pi r_1^2 r_5^2 \over 256 T^2_{\rm eff}}
   {\omega \left (\e^{\omega\over T_H} - 1 \right ) \over
   \left (\e^{\omega\over 2 T_L} - 1\right )
   \left (\e^{\omega\over 2 T_R} - 1\right ) }
   (\omega^2 + 16 \pi^2 T_L^2) (\omega^2 + 16 \pi^2 T_R^2) \ ,
  } 
 where $T_L$ and $T_R$ are the left and right-moving temperatures,
$T_{\rm eff}$ is the effective string tension
\refs{\LWW,\CTT,\ms,\juan,\tssm,\hrs} and $r_1^2$ and $r_5^2$ are
essentially the 1-brane and 5-brane charges.  The only restriction on
the validity of \SigmaDbrane\ is that $T_L,T_R,\omega \ll 1/{r_1} \sim
1/{r_5}$ so that we stay in the dilute gas regime and keep the
wavelength of the fixed scalar much larger than the longest length
scale of the black hole.  Remarkably, the very simple effective string
result \SigmaDbrane\ is in complete agreement with the rather
complicated calculations in semi-classical gravity!  The
semi-classical calculations involve no unknown parameters, so
comparison with \SigmaDbrane\ allows us to infer $T_{\rm eff}$.  The
result is in agreement with the fractional string tension necessary to
explain the entropy of near-extremal 5-branes \juan.

To set up the semi-classical calculations, we will develop in section
2 an effective action technique for deriving the equations of
motion for fixed scalars.  This technique shows how the fixed
scalar equation couples with Einstein's equations when $r_1 \neq r_5$;
therefore, we restrict ourselves to the regime $r_1= r_5 =R$ where
the fixed scalar equation is straightforward.
We briefly digress to four dimensions,
demonstrating how the same techniques lead to similar equations for
fixed scalars.  Clearly, comparisons analogous to the ones made in
this paper are possible for the four-dimensional case, where
the effective string appears at the triple intersection of M-theory
5-branes \KT.  In section 3 we use the Dirac-Born-Infeld (DBI) 
 action to see how various scalars in $D=5$ couple to the effective string.
The main result of section 3 is that the leading coupling of the
fixed scalar is to {\it four} fluctuation modes of the string.
This highlights its difference from the moduli which couple to two
fluctuation modes. In section
4  we return to five dimensions and exhibit approximate
solutions to the fixed scalar equation, deriving the semi-classical
emission and absorption rates. In section 5 we calculate the corresponding
rates with D-brane methods, finding complete agreement with semi-classical 
gravity. We conclude in section 6.
In the Appendix we discuss the absorption rate  as implied by the effective
string action of section 3 of some other  `off-diagonal' scalars
present in the system.

\newsec{Field Theory Effective Action Considerations}
\subsec{$D=5$ case}

First we shall concentrate on the case
of a $D=5$ black hole representing 
 the bound state of $n_1$ RR strings and  
$n_5$ RR 5-branes compactified on a 5-torus
\refs{\cm}. This black hole may be viewed as a  static solution 
corresponding to the following 
 truncation of type IIB superstring effective  action  
 compactified on 5-torus
   (cf. \refs{\maha,\bho})
 \eqn\actit{
S_5 = 
 {1 \ov 2 \k_5^2} \int d^5 x \sqrt{-g} \bigg[ R  - {4\ov 3}(\del_\m \p_5)^2 
 -  { 1 \ov 4} G^{pl} G^{qn}(\del_\m G_{pq} \del^\m G_{ln} 
 +  e^{{2} \p_5 } \sqrt {G} \del_\m B_{pq} \del^\m B_{ln} )
}  $$ -\ 
 \fourth e^{-{4\ov 3} \p_5   }G_{pq} F^{(K)p}_{\m\n}  F^{(K)q}_{\m\n} 
- \fourth e^{{2\ov 3}\p_5 } \sqrt {G} G^{pq}  H_{\m\n p } H_{\m\n q }
-  {1\ov 12} e^{{4\ov 3} \p_5 } \sqrt {G}  H^2_{\m\n\l}  
 \bigg]   \ ,  $$
where $\m,\n,...= 0,1,..,4; p,q,...=5,...,9$. \  
 $\p_5$ is the  5-d dilaton and $G_{pq}$ is the metric of
5-torus,  
$$\p_5 \equiv \p_{10} - \fourth \ln  G
\ , \qquad G= \det G_{pq}\  , $$ 
and $B_{pq}$ are the internal components of the  RR 2-form field.
 $F^{(K)p}_{\m\n}$ is the Kaluza-Klein
  vector field strength, while $H_{\m\n p}$ and $H_{\m\n\l}$
are  given explicitly by \maha\
  \eqn\dewf{
  H_{\m\n p} =   F_{\m\n p}  - B_{pq} F^{(K)q}_{\m\n} \ , 
  \ \ \ \  F_p = dA_p \ , \ \ \ F^{(K)p} = d A^{(K)p} \ , 
    }
  $$
  H_{\m\n\l} = \del_\m B'_{\n\l} - \hal  A^{(K)p}_\m F_{\n\l p}
  - \hal  A_{\m p} F^{(K)p}_{\n\l} + {\rm cycles} \ ,  $$
  where $A_{\m p}= B_{\m p}  + B_{pq} A^{(K)q}_\m $
  and $B'_{\m\n} = B_{\m\n}  + A^{(K)p}_{[\m} A_{\n] p} 
   -  A^{(K)p}_{\m} B_{pq} A^{(K)q}_{\n}  $
   are related to the  
 components  of the $D=10$
 RR  
 2-form field $B_{MN}$.  
 
  The  `shifts' in the field strengths in  \dewf\ 
  will vanish for the black hole 
 background
  considered  below (for which the internal components of the 
  2-form $B_{pq}$
 will be zero and the two 
  vector fields $A^{(K)p}$ and $A_p$ will be electric), and,
   as it turns out, 
 are  also  not  relevant for the discussion of perturbations.
 
  For comparison,  a similar truncated  $D=5$  action 
 with $B_{pq}$,  $F_{\m\n p}$ and $H_{\m\n\l}$  from the NS-NS  sector has 
  the following   antisymmetric tensor terms 
  (the full action in general 
  contains both RR and NS-NS antisymmetric tensor parts)
   \maha\
$$- { 1 \ov 4} G^{pl} G^{qn} \del_\m B_{pq} \del^\m B_{ln} 
- \fourth e^{-{4\ov 3}\p_5 }  G^{pq}  H_{\m\n p } H_{\m\n q }
-  {1\ov 12} e^{-{8\ov 3} \p_5 }   H^2_{\m\n\l} \ . $$

We shall assume that there are non-trivial electric charges in only 
one of the five  internal directions and  that the metric corresponding 
to  the internal 5-torus (over which the 5-brane will be wrapped) is 
\eqn\torr{
(ds^2_{10})_{T^5} = 
e^{2\n_5} dx_5^2 +  e^{2\n} 
(dx^2_6 + dx^2_7 + dx^2_8 + dx^2_9) \ ,  } 
where $x_5$ is the string direction  and   $\n$ is the `scale'
of the four 5-brane directions transverse to the string.
It is useful to introduce a different basis for the scalars, defining 
the `six-dimensional' dilaton,  $\p$, and the `scale' $\l$ of the  $x_5$ 
(string) direction as measured in the $D=6$ Einstein-frame metric:
\eqn\bass{
\p = \p_{10}-2\n= \p_5 + \ha \n_5 \ , \ \ \ \   
\l = \n_5 - \ha \p = {3\ov 4 }\n_5 - \ha \p_5 \ . }
The action  \actit\  can be expressed either in terms of
$\p_5, \n_5,\n$ or $\p,\l,\n$  (in both cases the kinetic term is
diagonal). In the latter case (we set $B_{pq}=0$) 
\eqn\acti{
S_5 = 
 {1 \ov 2 \k_5^2} \int d^5 x \sqrt{-g}
\bigg[ R  - (\del_\m \p)^2  - {4\ov 3} (\del_\m \l)^2  
 - 4  (\del_\m \n)^2 }  $$ -\ 
 \fourth e^{{8\ov 3}\l } {F^{(K)}_{\m\n}}^2 
- \fourth e^{-{4\ov 3} \l +  4\n } F_{\m\n}^2 
-  {1\ov 12} e^{{4\ov 3} \l + 4\n } H^2_{\m\n\l}  
 \bigg]   \ . $$
Here $F^{(K)}_{\m\n}\equiv F^{(K)5}_{\m\n}$ is the 
KK vector field strength corresponding 
to the string direction, while
$F_{\m\n}\equiv F_{\m\n 5}$ and $H_{\m\n\l}$ 
correspond  the `electric' (D1-brane) and `magnetic' (D5-brane)
components of the field strength of the  RR   2-form field.
Evidently $\p$ is an ordinary `decoupled' scalar while $\l$ and $\n$ 
are different: they interact with the gauge charges. 
We shall see that they are examples of the so-called `fixed scalars'.

To study spherically symmetric  
configurations corresponding to this action it is sufficient to 
 choose the five-dimensional metric in the  `2+3' form
 \eqn\met{
ds^2_5 = g_{mn} dx^mdx^n + ds^2_3=
 - e^{2a}dt^2 + e^{2b} dr^2   +  e^{2c}  d\Omega_3^2  \ ,  }
where $a,b,c$ are  functions of $r$ and $t$.
Solving first the equations for   
$H_{\m\n\l}$, $F_{\m\n}$  and $F^{(K)}_{\m\n}$
and assuming that  the first two have, respectively, 
the magnetic and the electric components
(with the  charges $P$ and $Q$ 
corresponding to the D5-brane   and the D1-brane), while
the third has only the electric component with the Kaluza-Klein 
charge $Q_K$, we may  eliminate them from the action \acti. 
The result is an effective  two-dimensional theory 
with coordinates $x^m=(t,r)$ and the action given (up to the
constant prefactor) by\foot{The full set of 
equations and constraints is derived by first keeping the 2-d metric $g_{mn}$
general and using its diagonal  gauge-fixed form only after the variation.  
In addition to choosing $g_{mn}$ diagonal as in \met, 
one  can  use the gauge freedom to  impose 
one more relation between $a $ and $b$.}
$$
S_2= \int d^2 x {\sqrt{-g}} e^{3 c } \bigg[R  +  6 (\del_m c)^2 
 - 
(\del_m \p )^2-  {4\ov 3}  (\del_m \l)^2 - 4(\del_m \n )^2 + V (c,\n,\l)  \bigg]   $$
 \eqn\ffoo{
= \int dtdr\bigg[ -  e^{3c + b -a} 
( 6\do c \do b + 6\do c^2 - \do \p^2 -  
{4\ov 3} \do \l^2    - 4  \do \n^2  ) } $$
+ \  e^{3c + a -b} ( 6c' a' + 6c'^2  -  \p'^2 -{4\ov 3}  \l'^2 - 4 \n'^2 )  
$$
$$+ \  6 e^{a+ b + c } - 2 e^{a+b-3c} f(\n,\l) \bigg] \ .   $$
The first term in the potential 
originates from the curvature of the 
3-sphere  while  the second is 
produced by the non-trivial charges, 
 \eqn\fffo{
 f(\n,\l) =     Q^2_K e^{-{8\ov 3}  \l}  +  e^{{4\ov 3}  \l} 
( P^2 e^{ 4\n } + Q^2 e^{-4 \n }) 
  \ .  }
 This is a special case of the more general expression following from  \actit: 
  if the electric charges corresponding to the vector 
fields in \actit\ are $Q_{Kp}$ and $Q^p$ we get 
\eqn\gene{f(\p_5, G_{pq})
 =  e^{{4\ov 3}  \p_5 }   Q_{Kp}  Q_{Kq} G^{pq}  
+  e^{-{2\ov 3}  \p_5} 
\bigg(    P^2   {G}^{1/2}  
+    Q^p Q^q  G_{pq} {G}^{-1/2}  \bigg) \ .}
The potential $f$ in \fffo\  has the  global minimum at 
$e^{ 4\n } = QP\inv, \ e^{{4}  \l} = Q^2_KQ\inv P\inv $.
These values of $\n$ and $\l$ are thus `fixed points'
to which these fields are attracted on the horizon, which is
why such fields can be called `fixed scalars.' By contrast, 
the decoupled scalar $\p$ can be chosen  to be equal to an 
arbitrary  constant.  

As an aside, we note that this  structure of the potential \fffo\ 
explains why one needs at least {\it three}  different charges
to get an extremal  $D=5$ black hole  with a regular horizon (i.e. 
with scalar fields that do not blow up): 
it is necessary to have 
  at least three exponential terms to `confine' 
  the  two fixed scalars. If the number of 
  non-vanishing charges is smaller than three, then
one or both scalars will blow up at the horizon.

Equivalent actions and potentials are found for theories that are
obtained from the one above by U-duality.
For example, 
in the case of the NS-NS  truncation of type II action,
which  has a $D=5$ black hole  solution representing 
a bound state of NS-NS strings and solitonic 5-branes, 
 we  can put the action in the form 
 \ffoo, where $\l$  is still the scale of the string direction
as measured by the 6-d metric, while the roles of 
$2\n$ (the scale of the 4-torus) 
and  $-\phi$ are interchanged.\foot{There exists an equivalent  representation
of this NS-NS action where the fixed scalars are 
the 5-d dilaton and 
the scale of the string direction, 
while the scale of the 4-torus is decoupled.}

In order to find the static black hole solution to \ffoo, we define
$\r =  2c +  a, $ $d\t = - 2 e^{-3c -a + b} dr$.
 Now \ffoo\ reduces to 
a  `particle'  action (we  choose  $\p=\const$)
\eqn\oqfw{
S_1=  \int d\t\bigg[  {3\ov 2 } 
(\del_\t \r)^2  - {3\ov 2}(\del_\t a)^2  -{4\ov 3} 
(\del_\t \l)^2 - 4 (\del_\t \n)^2  
+ \  {3\ov 2 } e^{ 2 \r } -  {1\ov 2 } e^{ 2a} f(\n,\l) \bigg] \ ,} 
which should be supplemented by the `zero-energy' constraint,
$$
 {3\ov 2 } 
(\del_\t \r)^2  - {3\ov 2}(\del_\t a)^2  -{4\ov 3} 
(\del_\t \l)^2 - 4 (\del_\t \n)^2  
-  {3\ov 2 } e^{ 2 \r } +  {1\ov 2 } e^{ 2a} f(\n,\l) =0 \ . $$
The special structure of $f$ in \fffo\  makes it possible to find a
simple analytic  solution of this  `Toda-type' system. 
Introducing new variables 
$\a= a - {4\ov 3} \l, \ \  \b= a  + {2\ov 3} \l +2 \n  , 
\   \g= a + {2\ov 3} \l -  2\n $
and using the special form \fffo\ of $f$, we can convert \oqfw\ 
to four non-interacting Liouville-like  systems (related  only 
through  the constraint) 
\eqn\fw{
S_1= \int d\t\bigg[  {3\ov 2 }(\del_\t \r)^2  - \ha (\del_\t \a)^2  -  
\ha (\del_\t \b)^2 -  \ha (\del_\t \g)^2 } 
$$+ \  {3\ov 2 } e^{ 2 \r } -  \ha  Q^2_K e^{2\a}  - \ha  P^2 e^{2\b}  
- \ha   Q^2 e^{2\g}  \bigg] \ .     $$

The general solution depends on the three gauge charges 
$P,Q,Q_K$ and one parameter which we will call $\mu$ which governs
the degree of non-extremality. In a convenient gauge, the solution reads 
\refs{\ATT,\cm,\CYY,\hms} 
\eqn\soo{ e^{2a} = h  {\H}^{-2/3}   \ , \ 
\ \ \  e^{2b} = h\inv  {\H}^{1/3} \ ,\ \ \ 
e^{2c} = r^2  {\H}^{1/3 }\  
 ,  \ \ \ {\H}\equiv H_\P H_\Q H_{\Q_K} \ , } 
\eqn\sool{ 
e^{2\l } = H_{\Q_K}  (H_\Q H_{\P})^{-{1\ov 2}}  \ ,
\ \  \ \ e^{4\n } =  H_{\Q} H_\P\inv  \  ,  
\ \ \  \ \ e^{2\p} =  
e^{2\p_{10,\infty}}\ ,   }  
$$ h = 1 - {2\m \ov r^2} \ , \ \ \  
H_\q = 1 + {\hat q \ov r^2} \ , \ \ \  \hat q \equiv  
\sqrt {q^2 + \m^2 } - \m \ ,  \ \ \   q=(P,Q,Q_K) \ .  $$
We have chosen the asymptotic values $\l_\infty$ and $\n_\infty$ 
to be zero. To compare with previous equations, 
we also note that $e^{2\r} = r^2(r^2-2 \m)$. 

In the  extremal limit, $\m=0$, one finds  
$$e^{-\a} = H_{Q_K},\qquad
e^{-\b} = H_P, \qquad  e^{-\g} = H_{Q}\ ,$$
where $H_q= c_q  + q\t$  and $\t=1/r^2$.
The constants  $c_{Q_K},c_P,c_Q$  must satisfy 
$ c_{Q_K} c_P c_Q =1 $ in order for the 5-d metric
to approach the Minkowski metric at  infinity.
The two remaining arbitrary constants 
correspond to the asymptotic values of $\l$ and $\n$.
As is clear from \ffoo,\fffo, shifting $\l$ and $\n$ by  constants
 is equivalent to a rescaling of $Q_K,Q,P$.
The assumption that 
$\n_{\infty} =0$ 
and 
$\l_\infty = \n_{5\infty} + \n_{\infty}  - \ha \p_{10,\infty} =0$ 
implies (setting $\a'=1$):
  \eqn\UnitChoice{
   V_4= e^{4\n_\infty} =1\ , \qquad
   {\cal R}^2 = e^{2\n_{5\infty}} = g = e^{\p_{10,\infty}}\ ,\ \ \ \ \
   \k_5^2 = {2\pi^2 g^2 \over {\cal R} V_4} \ , 
  }
 where $(2\pi)^4 V_4$ is the volume of $T^4$
in the $(6789)$ directions, while ${\cal R}$ is the radius of the
circle in direction $5$.  Then the `charges' $Q_K,Q,P$ are related to
the quantized charges $n_1,n_5,n_K$ as follows:
  \eqn\IntCharges{
   n_1 = {V_4 Q \over g} = {Q \over g} \ , \qquad 
   n_5 = {P \over g} \ , \qquad 
   n_K = {{\cal R}^2 V_4 Q_K \over g^2} = {Q_K \over g} \ .
  }
The somewhat unusual form of the last relation is due to our choice 
$\l_\infty=0$ instead of more standard  $\n_{5\infty}=0$.

In using the black hole solution \soo, \sool, we
will often find it convenient to work in terms of characteristic radii
rather than the charges, so we define
  \eqn\DefRadii{
   r_1^2 = \hat{Q} \ , \qquad
   r_5^2 = \hat{P} \ , \qquad
   r_K^2 = \hat{Q}_K \ , \qquad
   r_0^2 = 2 \mu \ .
  }
{}From the classical GR point of view, these parameters can take on 
any values. Recent experience has shown, however, that 
when the radii satisfy \mast
\eqn\constraint{ r_0, r_K \ll r_1, r_5 \  }
the black hole can be successfully matched to a bound state of D1-branes
and D5-branes (with no antibranes present) carrying a dilute gas of
massless excitations propagating along the bound 
D1-branes. Evidence 
for this gas can be seen directly in the energy, entropy and temperature
of the black hole solution. Introducing a new parameter $\sigma$ through
$$ 
r_K^2 = r_0^2 \sinh^2 \sigma
$$
one finds the following expressions \refs{\hms,\mast}
for the ADM mass, Hawking 
temperature and the entropy in the parameter region \constraint:
\eqn\ment{
M= { 2\pi^2\over \kappa_5^2} \left (r_1^2 + r_5^2+
\ha {r_0^2 \cosh 2\sigma} \right )\ ,
\ \ \
 \ \  T_H^{-1}={2\pi r_1 r_5\over r_0}\cosh\sigma \ , }
 $$ {\rm S}= {2\pi A_{\rm h} \over \k_5^2 } = 
 {4\pi^3 \over \kappa_5^2} r_1 r_5 r_0 \cosh\sigma
\ .$$
The entropy and energy are those of a gas of massless one-dimensional
particles with the left-movers and right-movers 
each having its own temperature \refs{\mast}:
\eqn\temps{ T_L = {r_0 \e^\sigma \over 2\pi r_1 r_5}\ ,
\qquad T_R = {r_0 \e^{-\sigma} \over 2\pi r_1 r_5}\ .
}
The Hawking temperature is related to these two temperatures by
  \eqn\THDef{
   {2\over T_H}={1\over T_L}+{1\over T_R} \ ,  }
a fact which also has a natural thermodynamic interpretation.
These results will be heavily used in later comparisons of classical 
GR results with D-brane calculations of corresponding quantities.


Let us now turn to the discussion of the propagation of perturbations on 
this black hole background. The goal will be to calculate the classical
absorption cross-section of various scalar fields and eventually to
compare them with comparable D-brane quantities. The behavior of `free'
scalars, like $\p$, is quite different from that of `fixed' scalars,
like $\l$ and $\n$. The spherically symmetric fluctuations of $\p$  
obey the standard  massless  Klein-Gordon equation in this background.
Namely, if $\delta \p = e^{i\o t} \td \p (r)$,  then 
\eqn\kgo{
 \left[r^{-3} {d\over dr} ( h r^3 {d\over dr} ) + \omega^2    
h\inv H_\P H_\Q H_{\Q_K}   \right] \td \p  =0 \ , }
This scattering problem, and its D-brane analog, have been analyzed at
length recently and we will have no more to say about it. The spherically
symmetric fluctuations of the metric functions $a,b,c$ and the scalars  
$\l,\n$ in general obey a complicated set of coupled 
differential equations.\foot{The spherically symmetric 
fluctuations of the gauge fields need 
not be considered explicitly: 
since the dependence on $H_{\m\n\l}$  and $F_{\m\n}$ is gaussian, they are 
automatically included when going from \acti\ to \ffoo.}
However, when the charges  $P$ and $Q$ are set 
equal, a dramatic simplification
occurs: the background value of  $\n$  in \sool\  (i.e. the `scale' of 
the transverse 4-torus) becomes  constant and its  small fluctuations 
$\delta\n$ decouple from those of the other fields.\foot{Similar 
simplification occurs when any two of the three charges are equal.
For example, if $P=Q_K$  we may introduce  
$\l' = -\ha (  \l -3 \n) , \ \n'= -\ha (\n + \l) $ 
(in terms of  which the kinetic part 
in the action \ffoo\ preserves its diagonal form) 
to discover that $\n'$ has decoupled fluctuations. The resulting equation
for $\delta \n'$ has the  same form as the equation for $\delta \n$ in
the case of $P=Q$.}
 The gaussian effective 
action for $\delta\n$ extracted from \ffoo\ is
\eqn\ffu{
\delta S_2= \int d^2 x {\sqrt{-g}} e^{3 c} \left[ - 4 (\del_m \delta  \n )^2
- 32 P^2 e^{-6 c +{ 4\ov 3} \l}  ( \delta  \n )^2  + ...  \right]  }
and spherically symmetric fluctuations $\delta\n=e^{i\o t}\td\n$ obey 
\eqn\kgoo{
 \left[r^{-3} {d\over dr} ( h r^3 {d\over dr} )
 + \omega^2    h\inv  H^2_\P H_{\Q_K} 
 - 8 P^2 r^{-6} H_\P^{-2}\right] \td \n = 0 \ . }
This is the standard Klein-Gordon equation \kgo\ augmented by
a space-dependent mass term originating from the expansion of the 
effective potential $f(\n,\l)$ in \fffo. This mass term falls off as $r^{-6}$
at large $r$, and, in the extremal case, blows up like $8/r^2$ near the
horizon at $r=0$. This is the $l(l+2)/r^2$ angular momentum barrier for
an $l=2$ partial wave in $D=5$. This `transmutation' of angular momentum
plays an important role in the behavior of the fixed scalar cross-section. 
For later analysis, it will be convenient to rewrite this equation
using the coordinate $\t= 1/r^2$: 
\eqn\kgee{
 \left[  { [ (1- 2\m \t){d\over d\t}]^2 } 
 + \fourth \omega^2   \t^{-3} ( 1+ \P\t)^2(1 + \Q_K\t) 
 - 2 {P^2  (1- 2\m \t) \ov  (1 + \P\t)^{2}} \right] \td \n = 0   \ . }

Remarkably, the extremal fixed scalar equation is {\it identical} to 
the equation for the fluctuations of the components of the antisymmetric 
tensor, $B_{ij}$, in the uncompactified spatial directions. 
Taking  $\m=0, \ i,j,k=1,2,3,4$,
and making the appropriate reduction
of \acti,\soo, we find
\eqn\beee{ \delta S_5  \sim \int dt  dr   r^3  e^{-a + {4\ov 3} \l + 4\nu} 
\big[  - (\del_t  \delta B_{ij})^2  
+ e^{3a} \del_k  \delta B_{ij} \del_k  \delta B_{ij}
  + ...  \big]\ .}  
Defining 
$$ \delta B_{ij}  = e^{-a - {2\ov 3} \l - 2\nu} C_{ij}
= H^{l/2}_P C_{ij}, \qquad l=2 \ ,$$
we obtain the following
equation for $C_{ij} (r,t) = e^{i\omega t} \tilde C_{ij} (r)$
at extremality:
\eqn\cij{
 \left[r^{-3} {d\over dr} (  r^3 {d\over dr} )
 + \omega^2    H_P H_Q H_{Q_K} 
 - 8 P^2 r^{-6} H_P^{-2}\right] \tilde C_{ij} = 0 \ .
}
Note that the mass term comes from
$$ H_P^{l/2} r^{-3} {d\over dr} (  r^3 {d\over dr} )
H_P^{-l/2} = l(l+2)P^2 r^{-6} 
H^{-2}_P= 8P^2 r^{-6} H^{-2}_P
$$
and turns out to be the same  as in  the extremal limit of \kgoo.
Had we started with a vector field in
$D=5$ we would instead have $l=1$ and the mass term would be 
$3P^2 r^{-6}H^{-2}_P$. We conjecture that the antisymmetric tensor
$B_{ij}$ is related to the fixed scalar by the residual supersymmetry
of the extremal black hole background. As we discuss in the next section,
a similar identity holds in $D=4$ between the fixed scalar and
the vector, $A_i$, equations.
\foot{This  raises the
question of why the supersymmetry explanation  applies to 
the scalar-tensor pair, but not to the scalar-vector one in $D=5$. 
  The answer presumably 
lies in the `electric' nature of the two vector fields in \acti.
In general, the equations for  spherically-symmetric
perturbations in a non-trivial
background need not be invariant under S-duality relating 
$B_{\mu\nu}$  and $A_\mu$ in $D=5$.}

Note that, when all the three  charges are equal, $P=Q=Q_K$, the background
value of the other scalar, $\lambda$, is constant as well.
Then the small fluctuations of this field decouple from gravitational
perturbations and satisfy the same equation as $\nu$, \kgoo.
If only two of the charges are equal, then there is
only one fixed scalar which has a constant background value 
and decouples from gravitational perturbations.
We would also like to know the fixed scalar scattering equations 
(and solutions) for the general 
$Q_K\neq Q\neq P$ black hole. This problem 
is surprisingly complicated due to mixing with
gravitational perturbations, and we have yet to solve it. 

To summarize, we have identified a set of scalars around the familiar
type II string $D=5$ black hole solution which merit the name of
`fixed scalars' in that their horizon values are fixed by the
background charges.  Their fluctuations in the black hole background
satisfy the Klein-Gordon equation, augmented by a position-dependent
mass term. In section 4 we will solve the new equations to find the
absorption cross-section by the black hole for these special scalars.

\subsec{$D=4$ case}

Previous experience  \refs{\mst,\myers,\KT,\ktI,\us,\gkgrey}  suggests 
that  one may be able to  extend the $D=5$ successes
in reproducing entropies and radiation rates with D-brane methods 
to $D=4$ black holes carrying 4 charges. Although we will not
pursue the $p$-brane approach to $D=4$ black hole dynamics in this
paper, this is a natural place to discuss $D=4$ fixed scalars 
and to record their scattering equations for later use. 

A convenient representation of the $D=4$ black hole with four different 
charges \refs{\CY,\CTT} is the $D=11$ supergravity configuration 
$5\bot5\bot5$ of three 5-branes intersecting over a common string 
\refs{\KT,\CT}. The three magnetic charges are related to the numbers 
of 5-branes in three different hyperplanes, while the electric charge 
has Kaluza-Klein origin. The  reduction to $D=4$ of the relevant
part  of the $D=11$ supergravity (or $D=10$ type IIA) action has the form
\eqn\actip{
S_4 = {1\ov 2 \k^2_4} 
\int d^4 x \sqrt{-g} \bigg[ R   - 2(\del_\m \n)^2 -
{3\ov 2}  (\del_\m \zeta)^2  - {4\ov 3} (\del_\m \xi )^2 - 
(\del_\m \eta )^2  }
$$
-  \fourth    e^{3 \zeta }  ( F^{(K)}_{\m\n})^2
 -  \fourth   e^{ \zeta }  \bigg( e^{-{8\ov 3} \xi }  ( F^{(1)}_{\m\n})^2 
+ e^{{4\ov 3}\xi} [ e^{2\eta }  ( F^{(2)}_{\m\n})^2
+   e^{-2\eta }  ( F^{(3)}_{\m\n})^2 ] \bigg)  \bigg] \ . $$
The scalar fields are  expressed in terms of components of the internal 
7-torus part of  the $D=11$ metric.  By the logic of the previous section,
the `scale' $\n$ of the 6-torus transverse to the intersection string is
a decoupled scalar, while the fields $\zeta,\xi,\eta$ 
(related to the scale of the string direction and the ratios of sizes 
of 2-tori shared by pairs of 5-branes) are fixed scalars.
If the  internal part of the $D=11$ metric   is 
$$ds^2_7=  e^{2\n_4} dx^2_{4} + 
e^{2\n_1} (dx_5^2 + dx_6^2) +e^{2\n_2} (dx_7^2 + dx_8^2) +
e^{2\n_3} (dx_9^2 + dx_{10}^2)\ ,$$
where $x_4$ is the direction of the common string, 
then 
$$\n=  \n_1 + \n_2 +\n_3 , \ \ 
\xi= \n_1 - \ha \n_2 - \ha \n_3, \  \ \eta= \n_3-\n_2,  \ \ 
\zeta= \n_4  + {2\ov 3} (\n_1 + \n_2 +\n_3) . $$

Using an ansatz for the 4-d metric similar to \met, 
solving for the vector fields, and 
substituting the result back into the action, 
we get the following effective two-dimensional action
(cf. \ffoo)
\eqn\ffe{
S_2= \int d^2 x {\sqrt{-g}} e^{2 c} \bigg[R  + 
 2 (\del_m c)^2 - 2(\del_\m \n)^2
 -  {3\ov 2}  (\del_m \zeta)^2  - {4\ov 3} (\del_m \xi )^2 - 
(\del_m \eta )^2  }
$$ + \ 2 e^{-2c} - \ha e^{-4 c } f(\zeta,\xi,\eta) \bigg] \  , $$
where 
\eqn\ffour{
f(\zeta,\eta,\xi)  =  
Q^2_K e^{-3 \zeta } +   e^{ \zeta } \left[ P_1^2 e^{-{8\ov 3} \xi }  
+ e^{{4\ov 3}\xi} (  P^2_2 e^{2\eta } +  P^2_3 e^{-2\eta })\right]
  \ .  }
As in the $D=5$ case, one finds that the special structure  of $f$
makes it possible to diagonalise the interaction term 
by a field redefinition and thus 
find  the static solution in a simple factorised form \refs{\CY,\CTT,\CT}
(cf. \soo) 
\eqn\sooo{
ds^2_4 = -e^{2a}dt^2 + e^{2b} dr^2 + e^{2c} d\Omega_2^3 \ ,  }
$$
e^{2a} = h  {\H}^{-1/2}   \ , \ \ \  e^{2b}= h\inv   {\H}^{1/2}  \ , 
\ \ \  e^{2c} = r^2  {\H}^{1/2 }\  
 ,  \ \ \ {\H}\equiv H_{\Q_K} H_{\P_1} H_{\P_2} H_{\P_3}\ ,  $$
\eqn\sooll{ 
e^{2\eta } =H_{\P_3}H_{\P_2}\inv\ , 
\ \
e^{2\xi } = H_{\P_1}(H_{\P_2}  H_{\P_3})^{-1/2}\ , 
\ \ 
e^{2\zeta } = H_{\Q_K}(H_{\P_1} H_{\P_2} H_{\P_3})^{-1/3}, }
$$ h = 1 - {2\m \ov r} \ , \ \ \  
H_\q = 1 + {\hat q \ov r} \ , \ \ \  \hat q 
\equiv  \sqrt {q^2 + \m^2 } - \m \ ,  
\ \ \   q=(Q_K,P_1,P_2,P_3) \ .  $$
As in the $D=5$ case, for the
 generic values of charges the spherically symmetric 
perturbations 
of this solution obey  a complicated 
system of equations (for  discussions of perturbations
of  
single-charged dilatonic black holes see, e.g., 
\refs{\pertuu}).
However, when the three magnetic charges are equal, $\eta$ and $\xi$
have constant background values, and so their 
 small spherically-symmetric fluctuations 
decouple from the metric perturbations, 
\eqn\ifr{
\delta S_2= \int d^2 x {\sqrt{-g}} e^{2 c}
 \left[ - (\del_m \delta \eta )^2
- 2 P^2 e^{-4 c  + \zeta + {4\ov 3} \xi } 
( \delta \eta )^2  + ...  \right]  \ ,  }
leading to the  following radial 
 Klein-Gordon equation with an extra mass term 
($\delta \eta (r,t) = e^{i\om t} \td \eta (r)$;  cf. \kgoo)
\eqn\goo{
 \left[r^{-2} {d\over dr} ( h r^2 {d\over dr} )
 + \omega^2    h\inv  H^3_\P H_{\Q_K} 
 - 2 P^2 r^{-4} H_\P^{-2}   \right] \td \eta = 0 \ . }
The same universal equation is found for $\delta \xi$.
In terms of  $\t= 1/r$ this becomes
\eqn\gee{
 \left[  { [ (1- 2\m \t){d\over d\t}]^2 } 
 +  \omega^2   \t^{-2} ( 1+ \P\t)^3(1 + \Q_K\t) 
 - 2 {P^2  (1- 2\m \t) \ov  (1 + \P\t)^{2}} \right] \td \eta = 0   \ . }
Represented in this form 
this is  very similar to \kgee\
found in the $D=5$ case: the differential operator and mass terms are
 exactly the same,  while the  frequency terms
are related by $\omega \to 2 \omega, \   \t^{-3}( 1+ \P\t)^2 \to  
\t^{-2}( 1+ \P\t)^3$.

In the extremal case   and  with 
 all four charges  chosen to be   equal, $Q_K=P$,
\goo\  reduces to the equation 
 studied in \kr.  The characteristic 
coefficient 2 in the mass term gives the effective potential of the form
$l(l+1)/r^2$ near the horizon, with $l=1$. Away from the horizon, the
fixed scalar equation differs from that of the $l=1$ partial wave
of the ordinary scalar.
Remarkably, however, in the extremal limit the 
fixed scalar equation \goo\ is {\it identical} to that
for the vector perturbations in the  extremal black hole background. 
This is true not only when all charges are equal 
(so that all scalars have constant background values)  but also 
in the  above case of $P_i=P \not=Q_K$.
Consider  perturbations $\delta A_i (r,t)$ ($i=1,2,3$;
  $ \delta A_0=0, \ \nabla_i A^i = \del_i A_i=0$)
of any of the three  `magnetic' vector fields  in \actip, 
\eqn\vee{
\delta S_4  \sim \int dt  dr   r^2  e^{\zeta} 
\bigg[  - e^{-2a} (\del_t  \delta A_i)^2  
+ e^{2a} \del_i  \delta A_j \del_i  \delta A_j
  + ...  \bigg] \ . }
Redefining the field to absorb the prefactor  and using  \sooll,  
 $\delta A_i =   e^{- a - \ha \zeta} C_i
= H^l_P   C_i $, $l=1$, we obtain 
the Klein-Gordon-type  equation  for $C_i(r,t)$ 
with an   extra mass term 
\eqn\uuu{ H_P^{l}  \Delta_3 H_P^{-l} 
=  l(l+1) P^2 
r^{-4} H_P^{-2}  = 2 P^2 r^{-4} H_P^{-2} 
 \ , } 
which is exactly the same as  in \goo\ in
 the $\m=0$  limit.

 This immediately implies that the absorption cross-section for the
fixed scalars should, in the extremal case, have the same soft
behavior $\sim \om^2$ (see \kr\ and below) as the vector 
cross-section \age.  The two cross-sections differ, however, in non-extremal
case.  Indeed, using methods similar to those in section 4, we find
(for $\omega\sim T_H$)
  \eqn\newcross{
   \sigma_{\rm abs}
     = A_{\rm h} \hat{P} \hat{Q}_K (\omega^2 + 4 \pi^2 T_H^2) \ ,
  }
 which no longer vanishes at $\omega=0$.  The facts presented above
are consistent with a possible explanation of the relation between the
coupled scalar and vector perturbations as being due to the residual
supersymmetry (the unbroken 1/8 of maximal supersymmetry \CY) present
in the black hole background in the extremal limit.

Finally, let us note that 
there exist other representations of
the  4-charge $D=4$ black hole.
For example
in the case of the  $2\bot2\bot5\bot5$ 
representation  \KT, or, equivalently,  the 
 U-dual  $D=4$ configuration in 
the NS-NS sector  with 
two  (electric and magnetic) charges coming from the  $D=10$ 
antisymmetric tensor
and two  (electric and magnetic) charges being of Kaluza-Klein  origin, 
we may parametrize the metric as
$$ds^2_{10} =  e^{2\n_4} dx^2_4 + e^{2\n_5} dx^2_5
+ e^{2\n}(dx^2_6 +  dx^2_7 + dx^2_8+ dx^2_9)\ . $$
This leads to 
the effective Lagrangian related to the above  one \ffe\
by a linear field redefinition and re-interpretation of the charges.
The  potential is (cf. \ffour)
$$f(\p,\n_4,\n_5) =e^{2\p} (  Q^2_K e^{2\n_4 } +  Q^2 e^{-2\n_4 })  
+ e^{-2\p} (  P^2_K e^{2\n_5 } +  P^2 e^{-2\n_5 }) \ ,
$$
where $\p$ is the 4-d dilaton. 
This shows that the scale  of the remaining 4-torus, $\n$, decouples.

\newsec{Effective String Couplings}

We now turn to a discussion of the effective action governing the 
absorption and emission of fixed scalars by the
bound state of D1- and D5-branes. 
We use the same framework as the recent demonstrations of agreement between 
GR and D-brane treatments of the absorption of generic decoupled scalars 
\refs{\cm,\dmw,\dm,\us}. We assume that: (i) the $D=5$ black hole 
is equivalent to $n_1$ D1-branes bound to $n_5$ D5-branes, with some
left-moving momentum;
(ii) that the low-energy dynamics of this system is described by the  
DBI action for a string with an effective tension $T_{\rm eff}$, 
and 
(iii) that the relevant bosonic oscillations of this effective string are 
only in the four 5-brane directions ($i=6,7,8,9$) transverse to the 1-brane.
These assumptions serve to specify the detailed couplings of external 
closed string fields, in particular the fixed scalars, to the D-brane degrees
of freedom. This is an essential input to any calculation of absorption
and emission rates and, as we shall see, brings fairly subtle features 
of the effective action into play. 

Specifically, we assume that the low-energy excitations of our system are
described by the standard D-string action 
\eqn\nam{
I=- T_{\rm eff} \int d^2 \s\ e^{-\p_{10}} \sqrt { - \det \g_{ab}}  + ... \ , 
\ \ \ \   \g_{ab} = G_{\m\n} (X) \del_a X^\m \del_b X^\n \ , }
where $ \p_{10}$ and $G_{\m\n}$ are the $D=10$ dilaton and string-frame 
metric. The specific dependence on $\p_{10}$ is motivated by the expected 
$1/g_{str}$ behavior of the D-string tension. The normalization constant 
of the tension, $T_{\rm eff}$, is subtle and will be discussed later. Our goal
is to
read off the couplings between excitations of the effective string and the
fluctuations of the metric and 
dilaton that correspond to the fixed scalars.

It should be noted that  the essential structure of  the effective string
 action  we are interested in
can be,  at least  qualitatively, understood using semi-classical 
effective field theory methods. A straightforward 
 generalization of  the extremal 
classical
solution \refs{\ATT,\cm} describing  a BPS bound state of a 
string and 5-brane  in which the string is localized on
the 5-brane\foot{Instead of talking about a bound state of several
single-charged D-strings and D5-branes with coinciding centers, 
it is sufficient, at the classical level, to 
consider just a single string and a single 5-brane having
charges $Q\sim n_1$ and $P\sim n_5$.}
 has the $D=10$ string metric ($m=1,2,3,4; i=6,7,8,9$)
$$ds^2_{10}=  H^{-1/2}_1  H^{-1/2}_5 (-dt^2 + dx_5^2) 
  + H^{1/2}_1  H^{1/2}_5 dx_m dx_m  
   +  H^{1/2}_1  H^{-1/2}_5 dx_i dx_i  \ , $$
   where $ H_5=H_5 (x_m) = 1 + P/x^2_m$ and 
   $H_1=H_1(x_m,x_i)$ is a solution of 
$$ [\del^m\del_m + H_5 (x_n) \del^i\del_i] H_1 =0$$ 
   such that for $P\to 0$ it approaches  the standard string
   harmonic function, 
   $H_1 \to  1  + Q/(x^2_i + x^2_m)^3$. If one averages 
   over the $x_i$-dependence
   of $H_1$ one  returns to the  original `delocalised'
   case,  $H_1= 1+Q/x^2_i$,  which  corresponds to the extremal limit
   of the $D=5$ black hole \soo,\sool\ with $Q_K=0$ (here    
   we  consider the `unboosted' string).
    The presence of the 5-brane breaks  the $O(1,1) \times O(8)$
 symmetry of the standard  RR  string
 solution   down to   $O(1,1) \times O(4) \times O(4)$.
 Since the  localized  solution also breaks 4+4 translational 
  invariances, the string soliton has   4+4
collective coordinates: $X^m(x_5,t), X^i(x_5,t)$.
 The  corresponding  $O(4) \times O(4)$ invariant 
 effective string action thus  
should have the  following  form in the static gauge,
$$ I = \int d^2\s  [ T_0 + T_\parallel \del^a X^i \del_a X^i 
+ T_\perp  \del^a X^m \del_a X^m + ... ]\  . $$ 
 The constants 
 $T_0, T_\parallel,  T_\perp$ 
can be  determined using standard 
methods (see, e.g.,  \khuri)  
by substituting the perturbed solution into the
$D=10$ effective field theory  action, etc.   
 $T_0$  is  proportional to the
ADM mass of the background, $T_0 \sim P+ Q$. The same should be 
 true  also for $T_\perp$, $T_\perp \sim P + Q$, 
  since $X^m$ describe  oscillations of
 the whole bound state in the common transverse 4-space.
At the same time,  $T_\parallel$ is the effective
 tension of the string  within the 5-brane, so that $ T_\parallel
 \sim Q$.
In the special cases $P=0$ and $Q=0$ these
  expressions  are in  obvious agreement with the standard
 results for  a  free  string  and a free  5-brane.
 In the case when $P \gg Q$, i.e. $n_5 \gg n_1$,
 we learn that  $T_\perp  \gg T_\parallel$,
 so that oscillations of the string in the four directions $X^m$
 transverse to the 5-brane can  be ignored.\foot{One may
  also give the following
 argument in support of the claim that transverse oscillations of the
string
 can be ignored when  the string is light compared to the 5-brane.
The classical action for a D-string  probe moving in the above
background produced by a bound state of R-R  string  and  5-brane
  has the following form:
$I_1 = T_0 \int d^2\sigma [e^{-\phi} \sqrt {-\det (G_{\m\n} +
B^{(NS)}_{\m\n})}
+  B^{(R)} + ... ]$ (we shall set the  world-sheet gauge field
to zero and choose the  static gauge). If the string is oriented
parallel to the $x^5$ direction of the 5-brane (a BPS
  configuration),
  the non-trivial part of the potential term cancels out \tsetl.
The same is true for the dependence on the 
 $H_1$-function in the second-derivative terms which have the form
  $I_1  = T_0 \int d^2\sigma[ 1
+ \ha \del^a X^i \del_a X^i + \ha H_5 (X) \del^a X^m \del_a X^m  +
...] $.
The function $H_5=1 + {P\ov X^2_m}$  thus   determines the metric
 of the `transverse'  part of the moduli
space (see also \kabat), i.e.
 it plays the role of an effective $T_\perp$
 which blows up when the string approaches the 5-brane.
 Thus the string probe  can  freely move within the 5-brane,
 but its transverse motions are suppressed.} 
If we further  assume, following \refs{\ms,\dmI},  that the string 
lying within the
5-brane has the effective 
length $L_{\rm eff} \sim Q P \sim n_1 n_5$,  
we finish with the following expression for the effective
string  tension  in the  relevant  directions parallel  to
the 5-brane: $T_{\rm eff}
 \sim T_\parallel/L_{\rm eff} \sim 
1/P\sim 1/n_5$.
 This picture is consistent with that suggested in \refs{\ms,\juan}
 and will
 pass a non-trivial test below.

  In accord with the assumption (iii), we thus drop terms involving
derivatives of $X^m= X^{1,2,3,4}$ (i.e. motions in the uncompactified
directions). We also eliminate two more string coordinates by choosing
the static gauge $ X^0= \s^0, \ X^5 = \s^1$ and write
  \eqn\namm{
   \g_{ab} \equiv \eta_{ab} + \hat\gamma_{ab} 
     = G_{ab}(x) + 2G_{i(a} (x)  \del_{b)} X^i  
                 + G_{ij} (x) \del_a X^i \del_b X^j \ . 
  }
 We make the Kaluza-Klein assumption that the background fields
$\p_{10}$ and $G_{\m\n}$ depend only on the external coordinates
$x^m\equiv X^m, m=0,1,...,5$. Since we are interested in linear
absorption and emission processes, we make a weak-field expansion in
powers of $\p_{10}$ and $h_{\mu\nu} \equiv G_{\mu\nu} -\eta_{\mu\nu}$, 
splitting $h_{ij}$ into traceless and trace parts: $h_{ij} = \bar h_{ij} +
\fourth \delta_{ij} h$.  Finally, we distinguish L and R string
excitations by introducing $\del_+ = \del_0 + \del_1 \ , \del_- = -
\del_0 + \del_1$. We can then use the formula
  $$
   \sqrt { - \det (\eta_{ab} + \hat\gamma_{ab})} 
     = 1 + \ha \hat\gamma_{+-} -  {1\ov 8}\hat\gamma_{++}\hat\gamma_{--}
         + {1\ov 16 }\hat\gamma_{+-}\hat\gamma_{++}\hat\gamma_{--} + \ldots
  $$
 to expand \nam, finding the following action for $X^i$:
\eqn\expa{
I_X = -T_{\rm eff} \int d^2 \s \bigg[1  
+  \ha  \del_+ X \del_- X   - {1\ov 8}  (\del_+ X)^2  (\del_- X)^2  + ... }
$$ 
+\  L_0 + L_1 + L_2 + L_3 + L_4 +  L'_4 + ...\bigg] \ , 
$$
\eqn\opo{ L_0 = \ha (h_{55} - h_{00})  - \p_{10} \ , \ \  \ \ 
 L_1 =  \ha h_{5i} (\del_+  + \del_-) X^i   \ ,   }
\eqn\ttt{
L_2 = - \ha \p   
  \del_+ X^i \del_- X^i + 
  \ha \bar h_{ij} \del_+ X^i \del_- X^j   
 - {1\ov 8} (h_{00} + h_{55}) [(\del_+ X)^2 + (\del_- X)^2] \ ,   }
\eqn\iit{
L_3 =   - {1\ov 4}  h_{5i} [\del_- X^i    (\del_+ X)^2  +
  \del_+ X^i   (\del_- X)^2]   \ , }
\eqn\iis{
L_4 =   {1\ov 8} (\p_{10} - \ha h)  (\del_+ X)^2  (\del_- X)^2 =
{1\ov 8} \p  (\del_+ X)^2  (\del_- X)^2
- {1\ov 4}  \nu   (\del_+ X)^2  (\del_- X)^2  \ , }
$$L'_4 =   {1\ov 16} (h_{55} +h_{00})
 \del_+ X \del_- X  [  (\del_+ X)^2  +
 (\del_- X)^2]   +  {1\ov 16} (h_{55} -  h_{00})
 (\del_+ X)^2  (\del_- X)^2
$$
\eqn\iise{
=  {1\ov 8 } \l [ \del_+ X \del_- X  (  (\del_+ X)^2  +
 (\del_- X)^2)   +   (\del_+ X)^2  (\del_- X)^2] }
$$
    +  \  {1\ov 16 } \p [ \del_+ X \del_- X  (  (\del_+ X)^2  +
 (\del_- X)^2)   +   (\del_+ X)^2  (\del_- X)^2] + O(h_{00}) \ . 
$$
The expansion has been organized in powers of derivatives of $X^i$
and we have kept terms at most linear in the external fields (since
we don't use them in what follows, we have dropped higher-order terms involving
$\bar h_{ij}$). We have also reorganized those fields in a way appropriate 
for the compactification on a five-torus:
\eqn\defs{
\n\equiv  {1\ov 8} h \ , \ \ \  
\p \equiv \p_{10} -  {1\ov 4} h \ , \ 
\ \ \ \l \equiv \ha h_{55}  + {1\ov 8} h  - \ha \p_{10}
 =  \ha h_{55} - \ha \p\  \ , \ \ \ 
 h\equiv h_{ii} \ ,    }
 where $\n$ is the scale of the 4-torus part of the 5-brane (if
$G_{ij} = e^{2 \nu} \delta_{ij}$ then, in the linearized
approximation, $h_{ii} = 8 \n $), $\p$ is the corresponding
six-dimensional dilaton and $\l$ is the scale of the fifth (string)
dimension measured in the six-dimensional Einstein metric. These are
the same three scalar fields that appear in the GR effective action
\acti, \ffoo.

  Since the kinetic terms in the effective action \acti, \ffoo\ are
diagonal in $\p$, $\nu$ and $\l$, we can immediately read off some
important conclusions from \ttt, \iis. The expansion in powers of
worldsheet derivatives is a low-energy expansion and, of the fields we
have kept, only the dilaton $\p$ is coupled at leading order.  It is
also easy to see that the `off-diagonal' components of the metric
$\bar h_{ij}$ have the same coupling as $\p$ to lowest order in
energy.  (These are the fields whose emission and absorption were
considered in \refs{\dmw,\dm}).  What is more interesting is that the
scalar $\n$ only couples at the next-to-leading order (fourth order in
derivatives). Note that its interaction term can be written in terms
of worldsheet energy-momentum tensors as $\n T^X_{++} T^X_{--}$.\foot{
There is a similar coupling for $\p$ which produces a subleading
correction to its emission rate.}  The scalar $\l$ likewise does not
get emitted at the leading order and does couple at the same order as
$\p$, but with a different vertex.\foot{The different vertices for
$\n$ and $\l$ probably reflect the different behavior of their
fluctuations (the non-decoupling of $\delta \l$ from metric
perturbations) in the case when $ Q_K$ is not equal to $Q=P$. } The
`graviton' components in the time and string directions $h_{00}$ and
$h_{55}$ couple to the string in a way similar to $\l$, which reflects
their mixing with $\l$ in the effective action \ffoo.  Indeed, the
vertex $(h_{00} + h_{55})(T^X_{++}+ T^X_{--})$ gives a vanishing
contribution to the amplitude of production of a closed string state:
it only couples a pair of left-movers or a pair of right-movers so
that the production is forbidden kinematically.  The important (and
non-trivial) point is that the simplest DBI action for the coupling of
the external fields to the D-brane gives the fields $\n$ and $\l$,
previously identified as the `fixed scalars', different (and weaker)
couplings to the effective string than the fields like $\p$ previously
identified as `decoupled scalars.' The precise couplings will
shortly be used to make precise calculations of absorption and
emission rates.

The action \expa\ is at best the bosonic part of a supersymmetric action.
In previous discussions of D-brane emission and absorption, it has been
possible to ignore the coupling of external fields to the massless
fermionic excitations of the D-brane. For the questions that interest us,
that will no longer be possible and we make
a specific proposal for the couplings 
of worldsheet fermions. In the successful D-brane description of the entropy
of rotating black holes \refs{\brek}, five-dimensional angular momentum 
is carried by the fermions alone. There are two worldsheet fermion doublets,
one right-moving and one left-moving. 
The $SO(4)$ rotations of the 
uncompactified $(1234)$ coordinates are 
decomposed as $SU(2)_L\times SU(2)_R$
and the obvious (and correct) choice is to take the left-moving
fermions, $S^a$, to be a doublet under $SU(2)_L$ and the right-moving 
fermions, $S^{\dot a}$ to be a doublet under $SU(2)_R$. This set of fermions 
may be bosonized as two  boson fields, $\varphi^1$ and $\varphi^2$. 
As mentioned above,
the next-to-leading 
terms in \iit,\iis\ can be written in terms of the $X$-field 
stress-energy tensor,
$$ 
T^X_{++} = (\partial_+ \vec X)^2\ , \qquad T^X_{--} = (\partial_- \vec X)^2\ .
$$
The obvious guess for the supersymmetric completion of these
interaction
terms is simply to add the bosonization fields $\vec\varphi$ to the
worldsheet energy-momentum tensors:
$$ 
T^X_{++} \rightarrow T^{\rm tot}_{++}= (\partial_+ \vec X)^2+ 
(\partial_+ \vec \varphi)^2 \ ,
$$
and similarly for $T^X_{--}$. This will have a crucial effect on the 
normalization of the fixed scalar absorption rate.

We also observe that  the  scalars $h_{5i}$ 
(`mixing'  the string direction with  the  four  transverse directions) 
which also couple to the gauge fields in the effective action approach
(see  \actit), 
have a yet different coupling to the effective string. The rates produced
by this coupling are calculated in the Appendix, with the 
conclusion that $h_{5i}$ is neither a fixed scalar of the kind
studied in \kr\ nor an ordinary scalar. Thus, the Nambu action
predicts the existence of a variety of massless scalar fields
which interact differently with the black hole. 


\newsec{Semi-Classical Description of Absorption}

  In this section we will mainly discuss the solution of the radial
differential equation one obtains for $s$-wave perturbations in the
fixed scalar $\nu$ related to the volume of the internal $T^4$ in
string metric \torr.  Let us start by restating some results of
section 2.  
 {}From \met, \soo\ and \sool\ one can read off the five-dimensional
Einstein metric:
  \eqn\MetAgain{
   d s_{5}^2 = -(H_{\hat{Q}} H_{\hat{P}} H_{\hat{Q}_K})^{-2/3} h d t^2 + 
                 (H_{\hat{Q}} H_{\hat{P}} H_{\hat{Q}_K})^{1/3} 
                  \left( {h\inv} d r^2 + r^2 d \Omega_{3}^2 \right)
 \ . }
To avoid the mixing between gravitational perturbations and the fixed
scalar, we will restrict 
ourselves to the case $P=Q$, i.e. $r_1 = r_5 = R$, $P^2= R^2 (R^2 +
r^2_0)$ (see \DefRadii). 
The small fluctuation equation, \kgoo, may be written as
\eqn\ExactNonExtreme{
   \left[ \left( h r^3 \partial_r \right)^2 +
      \left (r^2+ R^2 \right )^2 (r^2 +r_K^2) \omega^2 -    
      {8 r^4 R^4 \over (r^2+ R^2)^2}
h \left (1 + {r_0^2\over R^2}\right ) \right] \tilde\nu= 0 \ ,
  }
where $h = 1 - {r_0^2 \over r^2}$. Since we work in the regime
$r_0 \ll R$, we will neglect the last factor in the last term.

 Several different radial coordinates are useful in different regions.
The ones we will use most often are $u$ and $y$ defined by the relations
  \eqn\RadVarDef{
   1-\df{r_0^2}{r^2} = \e^{-r_0^2 / u^2} \ , \qquad
   y = \df{R^2 r_K \omega}{2 u^2} \ .
  }
 Note that $u \to 0$ and $y \to \infty$ at the horizon.

  The most efficient tool for obtaining the absorption cross-section
is the ratio of fluxes method used in \mast.  In all the cases
we will treat, the solution to \ExactNonExtreme\ whose near-horizon
limit represents purely infalling matter has the limiting forms
  \eqn\NearLim{
   \tilde\nu \approx \e^{\i y}
  }
 near the horizon and 
  \eqn\FarLim{
   \tilde\nu \approx \alpha \df{J_1(\omega r)}{\omega r}
    = \df{\alpha}{2} \df{H^{(1)}_1(\omega r) + H^{(2)}_1(\omega r)}{\omega r}
  }
 far from the black hole, where $J_1$
is the Bessel function.\foot{In fact there can be phase shifts in the
arguments of the exponential in \NearLim\ and the Bessel functions in
\FarLim, but they are immaterial for computing fluxes.}  The term in
\FarLim\ containing $H^{(2)}_1(\omega r)$ is the incoming wave.  Once
the constant $\alpha$ is known, one can compute the flux for the
incoming wave and compare it to the flux for the infalling wave
\NearLim\ to find the absorption probability.  In the present
instance, fluxes are purely radial:
  \eqn\FluxOneForm{
   J = \df{1}{2\i} (\tilde\nu^* d \tilde\nu- {\rm c.c.}) 
     = J_r d r \ .
  }
 Observing that the number of particles passing through a sphere $S^3_r$
at radius $r$ in a time interval $[0,t]$ is 
  \eqn\NumPass{
   \int_{S^3_r \times [0,t]} *J = 2 \pi^2 h r^3 J_r t \ ,
  }
 one concludes that the flux per unit solid angle is 
  \eqn\Flux{
   {\cal F} = h r^3 J_r 
            = \df{1}{2\i} (\tilde\nu^* h r^3 \partial_r \tilde\nu - 
                {\rm c.c.}) \ .
  }
 The absorption probability is
  \eqn\AbsProb{
   1 - |S_0|^2 = {{\cal F}_{\rm h} \ov  {\cal F}_\infty^{\rm incoming}} 
    = \df{2 \pi}{|\alpha|^2} R^2 r_K \omega^3 \ .
  }
 We will always be interested in cases where this probability is small.
 By the Optical Theorem, the absorption cross-section is
  \eqn\SigmaAbs{
   \sigma_{\rm abs} = \df{4 \pi}{\omega^3} \left( 1 - |S_0|^2 \right)
     = \df{8 \pi^2}{|\alpha|^2} R^2 r_K .
  }
 Readers unfamiliar with the solution matching technology may be
helped by the analogy to tunneling through a square potential barrier
in one dimension.  If particles come from the left side of the
barrier, the wave function is to a good approximation a standing wave
on the left side of the barrier, a decreasing exponential inside the 
barrier, and a purely right moving exponential on the right side of 
the barrier.

  To obtain the familiar result $\sigma_{\rm abs} = A_{\rm h}$ for
low-energy, ordinary scalars falling into an extremal black hole, it
suffices to match the limiting value of \NearLim\ for small $y$
directly to the limiting value of \FarLim\ 
for small $r$ \dgm.  Due to
non-extremality and to the presence of the potential term in
\ExactNonExtreme, this naive matching scheme is invalid.  A more
refined approximate solution must be used, and a more physically
interesting low-energy cross-section will be obtained.  

  We will now present approximate solutions to \ExactNonExtreme\ in two
regimes most easily characterized in D-brane terms: we shall first
consider $T_R = 0$ with $\omega/T_L$ arbitrary; then we shall consider
$T_R$ much less than $T_L$ but not zero, and allow $\omega/T_R$ to
vary arbitrarily.  

  When $T_R = 0$, the black hole is extremal: $r_0 = 0$ and $r = u$.
As usual,  one proceeds by joining a near horizon solution ${\bf I}$ to
a far solution ${\bf III}$ using an exact solution ${\bf II}$ to the
$\omega = 0$ equation \refs{\dm,\kr}.  
The dominant terms of \ExactNonExtreme\ and 
the approximate solutions in the three regions are 
  \eqn\EtaSols{\vcenter{\openup1\jot
   \halign{\strut\span\TL & \span\TR & \span\TT & 
                 \span\TL & \span\TR\cr
   {\bf I.}\ \ &\left[ \partial_y^2 + 1 - 
      \df{2\eta}{y} - \df{2}{y^2} \right] \tilde\nu_{\bf I} = 0 &
    \qquad\qquad & \tilde\nu_{\bf I} &= G_1(y) + \i F_1(y) \cr
   {\bf II.} \ \ &\left[ (r^3 \partial_r)^2 - 
      8 \df{R^4}{H^2} \right] \tilde\nu_{\bf II} = 0 &
    \qquad\qquad & \tilde\nu_{\bf II} &= {C \over H(r)} + D H^2 (r) \cr
   {\bf III.} \ \ &\left[ (r^3 \partial_r)^2 + 
      r^6 \omega^2 \right] \tilde\nu_{\bf III} = 0 &
    \qquad\qquad &
     \tilde\nu_{\bf III} &= \alpha \df{J_1(\omega r)}{\omega r} + 
      \beta \df{N_1(\omega r)}{\omega r} \ , \cr
  }}}
 where $C,D,\a$ and $\b$ are constants, $H = 1 + R^2/r^2$, and $F_1$ and
$G_1$ are Coulomb functions \hmf\ whose charge parameter $\eta$ is
given by
  \eqn\EtaDef{
   \eta = -\tf{1}{4} \left( 2 \omega r_K + \df{\omega R^2}{r_K} \right) 
        = -{1 \over 4 \pi} {\omega \over T_L} 
            \left( 1 + 2 {r_K^2 \over R^2} \right) \ .
  }
 In the last equality we have used the definition of $T_L$ in \temps.
The quantity $r_K^2 / R^2$ is small in the dilute gas approximation,
and we will neglect it when comparing the final semi-classical 
cross-section with the D-brane answer.

  By design, $\tilde\nu_{\bf I} \to \e^{\i y}$ as $y \to \infty$
up to a phase shift in $y$.  An approximate solution can be patched
together from $\tilde\nu_{\bf I,II,III}$ if one sets
  \eqn\EtaMatch{
    C = \df{\alpha}{2} = \df{2}{3 C_1(\eta) \omega r_K} \ , \qquad 
    D = 0 \ , \qquad
    \beta = 0 \ ,
  }
where $C_1(\eta) = { 1 \over 3} e^{-\pi \eta / 2} |\Gamma(2+i\eta)| $ \hmf.
 A slightly better matching can be obtained by allowing $D$ and
$\beta$ to be nonzero, but the changes in the final solution do not
affect the fluxes ${\cal F}_{\rm h}$ and ${\cal F}_\infty$ (these
changes are however crucial in determining $S_0$ by the old methods of
\Unruh, and give phase information on the scattered wave which the
flux method does not).  Having only $C \neq 0$ in region $\bf II$ is
analogous to the fact that for right-moving particles incident on a
square potential barrier, the wave function under the barrier can be
taken as a purely falling exponential with no admixture of the rising
exponential.

  {}From \EtaMatch\ and \SigmaAbs\ the cross-section is immediate:
  \eqn\EtaSigma{
   \sigma_{\rm abs} = {\pi^2 \over 2} r_K R^2
 (\omega r_K)^2 |3 C_1(\eta)|^2
     = \tf{1}{4} A_{\rm h} (\omega r_K)^2 (1 + \eta^2) 
        \df{2 \pi \eta}{\e^{2 \pi \eta} - 1} \ ,
  }
where $A_{\rm h}$ is the area of the horizon (given in  \ment).
Note that the derivation of \EtaSigma\ does not require 
the assumption that $r_K \ll R$.

  To make the  comparison with  
  the D-brane approach, we  can 
  write \EtaSigma\ in the following 
suggestive form
  \eqn\ESAgain{
   \sigma_{\rm abs} = {\pi^2 \over 2} r_K R^2 (\omega r_K)^2 
    {{\omega \over 2 T_L} \over 1 - \e^{-{\omega \over 2 T_L}}} 
\left (1 + {\omega^2 \over 16 \pi^2 T_L^2} \right )
     \left[ 1 + O(r_K^2/R^2) \right] \ .
  }
 In section 5 we will compute the same quantity using effective
D-string method and will find agreement to the indicated order of
accuracy.  To obtain $O(r_K^2/R^2)$ corrections on the D-brane side
one would have to go beyond the dilute gas approximation.  
An interesting special case where these corrections vanish is when
$T_L = 0$, corresponding to $\eta \to -\infty$.  In the brane
description, this corresponds to 1-branes and 5-branes only with no
condensate of open strings running between them: a pure quantum state
with no thermal averaging.  The limiting forms of the GR result
\EtaSigma\ and of the D-brane absorption cross-section \SigmaDbrane\
to be derived in section~5 then agree exactly:
  \eqn\SigmaPure{
   \sigma_{\rm abs} = \left( \tf{\pi}{4} \right)^3
    R^8 \omega^5 \ .
  }

  Now let us continue on to the second regime in which an approximate
solution to the radial equation \ExactNonExtreme\ is fairly
straightforward to obtain: $\omega,T_R \ll T_L$ with $\omega/T_R$ 
arbitrary.  A quantity which enters more naturally into the 
differential equations than $\omega / T_R$ is 
  \eqn\ADef{
   B = {R^2 r_K \omega \over  r_0^2  }  = {\omega \ov \kappa } \tanh 
   \sigma\approx 
   {\omega \ov \kappa } \  , 
    \ \ \ \  \k\equiv 2\pi T_H \
   , 
  }
 where  $\k$ is the surface gravity at  the horizon, 
  and in the last step we used the fact that $\sigma \gg 1$ when
$T_R\ll T_L$.  In dropping terms from the exact equation
\ExactNonExtreme\ to obtain soluble forms in the three matching
regions, it is essential to retain $B$ as a quantity of $O(1)$;
however, $r_0/r_K$ and $\omega R^2 / r_K$ can be regarded as small
because $T_R \ll T_L$ and $\omega \ll T_L$.  In regions ${\bf II}$ and
${\bf III}$, the approximate equations turn out to be precisely the
same as in \EtaSols, but in ${\bf I}$ one obtains a more complicated
differential equation:
  \eqn\NESolI{
\left[ \partial_y^2 + 1 - 
      { 8\ov B^2} {\e^{-2y/B} \over (1 - \e^{-2y/B})^2} \right] 
       \tilde\nu_{\bf I} = 0  \ .
  }
 This equation can be cast in the form of a supersymmetric quantum
mechanics eigenfunction problem.  Define a rescaled variable $z =
y/B$ and supercharge operators
  \eqn\SuperQ{
   \QQ = -\partial_z + {\rm coth}\, z \ , \qquad
   \QQ^\dagger = \partial_z + {\rm coth}\, z \ .
  }
 Then \NESolI\ can be rewritten in the form
  \eqn\HOne{
   \QQ \QQ^\dagger \tilde\nu_{\bf I}
     = \left[ -\partial_z^2 + 2 {\rm csch}^2\, z + 1 \right] 
        \tilde\nu_{\bf I}
     = \left( 1  + B^2 \right) \tilde\nu_{\bf I} \ .
  }
 The eigenfunctions of the related Hamiltonian $\QQ^\dagger \QQ =
-\partial_z^2 + 1$ are just exponentials, and from them one can read
off the solutions to \HOne: the infalling solution is
  \eqn\Infall{
   \tilde\nu_{\bf I} 
     = {\QQ \, \e^{\i Bz} \over 1-\i B } 
     = {{\rm coth}\, z - \i B \over  1 -\i B} \,
         \e^{\i B z} 
     = { {\rm coth}\, {y\ov B} -\i B \over
         1 -\i B } \, \e^{\i y} \ .
  }
 The factor in the denominator is chosen so that $\tilde\nu_{\bf
I} \approx \e^{\i y}$ for large $y$.

  Performing the matching as usual, one obtains
  \eqn\Interp{
   \sigma_{\rm abs} = \tf{1}{4} A_{\rm h} (\omega r_K)^2 (1 + B^{-2} )
=\tf{1}{4} A_{\rm h} (\omega r_K)^2 \left (1+ {4 \pi^2 T_H^2\over
\omega^2 }\right )\ .
  }
In section~5 we will  show that the effective string
calculation gives the same result when
$\omega,T_R \ll T_L$.


\newsec{D-brane Absorption and Emission Cross-Sections}

In this section we give a detailed calculation of the emission and 
absorption of the fixed scalar $\n$, using the interaction vertices
computed in section 3. We recall that  $\nu$ is related (see \torr) 
to the
  volume   (measured in the string metric) 
  of the compactification 4-torus
 orthogonal to the string. To study the leading coupling
of $\nu$, it is sufficient to retain the following two terms in
the string effective action (cf. \expa):
  \eqn\Retain{\eqalign{
    I &= \int d^2 \sigma \bigg\{ -{1\over 2} (\partial_+ \vec X\cdot
         \partial_- \vec X + \partial_+ \vec\varphi \cdot 
          \partial_- \vec\varphi) \cr
      &\qquad\ + {1\over 4 T_{\rm eff}} 
\left [(\partial_+ \vec X)^2+ (\partial_+ \vec \varphi)^2\right ]
\left [(\partial_- \vec X)^2+ (\partial_- \vec \varphi)^2\right]
 \ \nu (x)  \bigg\}
  }}
where we have absorbed $\sqrt {T_{\rm eff}}$ into the fields
to make them properly normalized.  {}From \acti\ we see that the
scalar field with the proper bulk kinetic term is $2 \nu/\kappa_5$.  
Consider the invariant amplitude for processes mediated by the quartic 
interaction in \Retain. If $p_1$ and $p_2$ are the left-moving energies, while
$q_1$ and $q_2$ are the right-moving ones, the matrix element among
properly normalized states is
  \eqn\MatrixElem{
   {\sqrt 2\kappa_5\over T_{\rm eff}} \sqrt{ q_1 q_2 p_1 p_2\over \omega} \ .
  }

The basic assumption of the D-brane approach to black hole physics
is that the left-movers and right-movers can be treated as thermal
ensembles \refs{\cm,\hms}. 
Strictly speaking, they are microcanonical 
ensembles, but for our purposes the canonical ensemble is good enough
and we proceed as if we are dealing with a massless one-dimensional
gas of left-movers of temperature $T_L$ and right-movers with temperature
$T_R$. The motivation for this assumption has been explained at length
in several  recent papers \refs{\cm,\hms,\mast}.
To compute the rate for the process $scalar \to L+L+R+R$
we have to square the normalized matrix element \MatrixElem\ and
integrate it over the possible energies of the final state particles.
Because of the presence of the thermal sea of left-movers and right-movers, 
we must insert Bose enhancement factors: for example, each left-mover in 
the final state picks up a factor of $1 + \rho_L(p_i) = -\rho_L(-p_i)$, where
  \eqn\BEDist{
   \rho_L (p_i) = {1\over  \e^{p_i\over T_L}-1 }
  }
is the Bose-Einstein distribution.  If there were a left-mover of
energy $p_i$ in the initial state, it would pick up an enhancement
factor $\rho_L(p_i)$.  Similar factors attach to right-movers.

Conservation of energy and of momentum parallel to the effective
string introduces the factor
  \eqn\DeltaFn{
   (2\pi)^2 \delta(p_1+ p_2 + q_1 + q_2-\omega)
    \delta(p_1+ p_2- q_1 -q_2) = {1\over 2} (2\pi)^2 \delta(p_1+ p_2-
    {\omega\over 2}) \delta(q_1 + q_2 -{\omega\over 2})
  }
into the integrals over $p_1$, $p_2$, $q_1$, and $q_2$.
Putting everything together, we find that the rate for $scalar \to
L+L+R+R$ is given by
  \eqn\firstrate{\eqalign{
    \Gamma(1) &= \Gamma(scalar \to L+L+R+R) \cr
     &= {36\over 4} {\kappa_5^2 L_{\rm eff} \over 4 \pi^2 T^2_{\rm eff}\omega } 
         \int_0^\infty d p_1 d p_2 \, 
          \delta\left( p_1 + p_2 -{\omega\over 2} \right)
          {p_1 \over 1 - \e^{-{p_1\over T_L}}} 
          {p_2 \over 1 - \e^{-{p_2\over T_L}}} \cr
    & \qquad\qquad \times 
         \int_0^\infty d q_1 d q_2 \, 
          \delta\left( q_1 + q_2 -{\omega\over 2} \right)
          {q_1 \over 1 - \e^{-{q_1\over T_R}}}
          {q_2 \over 1 - \e^{-{q_2\over T_R}}} \ , 
  }}
where $L_{\rm eff}$ is the length of the effective string.
The factor of $36=6^2$ arises from the presence of 
six species of left-movers (four bosons and two bosonized fermions)
and six species of right-movers. We divide by $4 = 2^2$ because of
particle identity: because the two left-movers in the final state are
identical particles, the integral over $p_1,p_2$ counts every
left-moving final state twice (similarly for the right-movers).

  To write down the rates for the three other absorptions processes
(that is, processes $2$, $3$, and $4$  in   eq. \FixedDProc), it is
convenient to define the integrals
  \eqn\LRInt{\eqalign{
   I_L(s_1,s_2) &= \int_0^\infty d p_1 d p_2 \,
          \delta\left( s_1 p_1 + s_2 p_2 + {\omega \over 2} \right)
          s_1 p_1 \rho_L(s_1 p_1) \cdot s_2 p_2 \rho_L(s_2 p_2) \cr
   I_R(s_1,s_2) &= \int_0^\infty d q_1 d q_2 \,
          \delta\left( s_1 q_1 + s_2 q_2 + {\omega \over 2} \right)
          s_1 q_1 \rho_L(s_1 q_1) \cdot s_2 q_2 \rho_L(s_2 q_2) \ .
  }}
 The choices $s_i = 1$ and $s_i = -1$ correspond, respectively, to 
putting a particle in the initial or final state.  Then the total 
absorption rate, including all four competing processes of \FixedDProc,
is
  \eqn\AbsorbAll{\eqalign{
   \Gamma_{\rm abs}(\omega) 
    &= \Gamma(1) + \Gamma(2) + \Gamma(3) + \Gamma(4) \cr
    &= 36 {\kappa_5^2 L_{\rm eff} \over 4 \pi^2 T_{\rm eff}^2 \omega}
        \Big[ \tf{1}{4} I_L(-1,-1) I_R(-1,-1) + 
              \tf{1}{2} I_L(-1, 1) I_R(-1,-1) \cr
    & \qquad\qquad + 
              \tf{1}{2} I_L(-1,-1) I_R(-1, 1) + 
                        I_L(-1, 1) I_R(-1, 1) \Big] \cr 
    &= {9\kappa_5^2 L_{\rm eff} \over 4\pi^2 T_{\rm eff}^2 \omega}
       \int_{-\infty}^\infty d p_1 d p_2 \, 
        \delta\left( p_1 + p_2 -{\omega\over 2} \right)
        {p_1 \over 1 - \e^{-{p_1\over T_L}}} 
        {p_2 \over 1 - \e^{-{p_2\over T_L}}} \cr
    & \qquad\qquad \times 
       \int_{-\infty}^\infty d q_1 d q_2 \, 
        \delta\left( q_1 + q_2 -{\omega\over 2} \right)
        {q_1 \over 1 - \e^{-{q_1\over T_R}}}
        {q_2 \over 1 - \e^{-{q_2\over T_R}}} \cr
    &= {\kappa_5^2 L_{\rm eff} \over (32 \pi)^2 T_{\rm eff}^2}
        {\omega \over 
         \left(1- \e^{-{\omega\over 2 T_L}} \right)
         \left(1- \e^{-{\omega\over 2 T_R}} \right) 
        }
        \left( \omega^2 + 16 \pi^2 T_L^2 \right)
        \left( \omega^2 + 16 \pi^2 T_R^2 \right) \ .
  }}
 The fractional coefficients inside the square brackets on the second
line of \AbsorbAll\ are symmetry factors for the final states (the
initial states are always simple enough so that their symmetry factors
are unity).  It is remarkable that although the individual processes
$1$--$4$ have rates which cannot be expressed in closed form, their
sum is expressible in terms of integrals which can be performed
analytically \GR\ because they run over all $p_1$, $p_2$, $q_1$,
and~$q_2$.

  A similar calculation may be performed for the four emission processes, 
with the result
\eqn\EmitAll{
   \Gamma_{\rm emit}(\omega) 
    = \e^{-{\omega \over T_H}} \Gamma_{\rm abs}(\omega) 
    = -\Gamma_{\rm abs}(-\omega) \ ,
  }
where the Hawking temperature characterizing the distribution of the
emitted scalars is related to $T_R$ and $T_L$ by \THDef.
Our convention has been to compute $\Gamma_{\rm abs}(\omega)$
assuming that the flux of the incoming scalar is unity.  We have also
suppressed the phase space factor $\d^4 k / (2 \pi)^4$ for the
outgoing scalar in computing $\Gamma_{\rm emit}(\omega)$, and we have
assumed that the outgoing scalar is emitted into the vacuum state, so
that $\Gamma_{\rm emit}(\omega)$ includes no Bose enhancement factors.
These conventions were chosen because they lead to simple expressions
for $\Gamma_{\rm emit}(\omega)$ 
\EmitAll\ and $\sigma_{\rm abs}$  below,
 but they must be borne carefully in mind
when considering questions of detailed balance.  Suppose we put the
black hole in a thermal bath of scalars at temperature $T_H$.  Then
$\Gamma_{\rm abs}(\omega)$ and $\Gamma_{\rm emit}(\omega)$ pick up
Bose enhancement factors for the scalars: those factors are,
respectively, $1/(\e^{\omega/T_H}-1)$ and $1/(1-\e^{-\omega/T_H})$.
Once these factors are included, the emission and absorption rates
become equal by virtue of the first equality in \EmitAll.  The
fact that calculating $\Gamma_{\rm emit}(\omega)$ in the same way that
we calculated $\Gamma_{\rm abs}(\omega)$ leads to \EmitAll\ is a
nontrivial check on detailed balance.  This check is analogous to 
verifying that QED reproduces the Einstein $A$ and $B$ coefficients
for the decay of the first excited state of hydrogen.

  Because $\Gamma_{\rm abs}(\omega)$ was computed assuming unit flux,
one would naively guess that the absorption cross-section to be
compared with a semi-classical calculation is $\sigma_{\rm abs} =
\Gamma_{\rm abs}(\omega)$.  (Now we are back to the conventions where
$\Gamma_{\rm abs}(\omega)$ and $\Gamma_{\rm emit}(\omega)$ do not
include Bose enhancement factors for the scalars).  This is not quite
right; instead,
  \eqn\SigmaGamma{
   \sigma_{\rm abs}(\omega) 
     = \Gamma_{\rm abs}(\omega) - \Gamma_{\rm emit}(\omega) 
     = \Gamma_{\rm abs}(\omega) + \Gamma_{\rm abs}(-\omega) \ .
  }
 To see why \SigmaGamma\ is right, we have to remember what we are
doing in a semi-classical computation.  We send in a classical wave in
the field whose quanta are the scalars of interest, and we watch to
see what fraction of it is sucked up by the black hole and what
fraction is re-emitted.  The quantum field theory analog is to send
in a coherent state of scalars with large average particle number, so
that the flux is almost fixed at its classical expectation value $\cal
{}F$.  The dominant processes are then absorption and stimulated
emission.  The Bose enhancement factors collapse to $\cal {}F$ for
both absorption and emission, up to errors which are insignificant in
the semi-classical limit.  The net rate at which particles are
absorbed is then $\Gamma_{\rm abs}(\omega) {\cal {}F} - \Gamma_{\rm
emit}(\omega) {\cal {}F}$.  But this rate is $\sigma_{\rm abs} {\cal 
{}F}$ by definition, whence \SigmaGamma.  Note that the last expression
in \SigmaGamma\ is manifestly invariant under time-reversal, which takes
$\omega \to -\omega$.

In order to obtain definite results for the absorption cross-section, we
must supply values for the effective length $L_{\rm eff}$ of the string, as
well as its effective tension $T_{\rm eff}$. It is a by-now-familiar story
that multiple D-strings bound to multiple five-branes behave like a
single D-string multiply wound about the compactification 
direction \ms.
In the case at hand it is well-understood that the effective string
length is \refs{\ms,\mast}
\eqn\klrel{\kappa_5^2 L_{\rm eff} = 4 \pi^3 r_1^2 r_5^2 \ .}
With this substitution, 
the fixed scalar $\n$ absorption cross-section  becomes
\eqn\SigmaD{
\sigma_{\rm abs}(\omega) = { \pi r_1^2 r_5^2 \over 256 T^2_{\rm eff}}
{\omega \left (\e^{\omega\over T_H} - 1 \right ) \over   
\left (\e^{\omega\over 2 T_L} - 1\right )
\left (\e^{\omega\over 2 T_R} - 1\right ) }
(\omega^2 + 16 \pi^2 T_L^2) (\omega^2 + 16 \pi^2 T_R^2)\ .
}
This is similar to, but not quite the same as, the absorption cross-section
for the ordinary `unfixed' scalar calculated in \mast.

The object of our exercise is to offer further evidence that the
D-brane configuration {\it is} the corresponding black hole by showing
that \SigmaD\ is identical to the corresponding quantity calculated
by standard classical GR methods. For technical reasons, the GR calculation
in a general black hole background is quite difficult and the results we
have been able to obtain (presented in section 4) are only valid 
in certain simplifying limits. The most important simplification is to
take equal brane charges $r_1=r_5=R$. 

First we consider the extremal limit, $T_R=0$. Here \SigmaD\ 
reduces to
\eqn\simpler{
 \sigma_{\rm abs}(\omega) = { \pi^3 r_1^2 r_5^2 T_L^3 \over 8 T^2_{\rm eff}}
\omega^2 {{\omega \over 2 T_L} \over 1 - \e^{-{\omega \over 2 T_L}}} 
\left (1 + {\omega^2 \over 16 \pi^2 T_L^2} \right )\ . }
This is to be compared with the classical fixed scalar absorption
cross-section in the extremal background (eqn. \ESAgain):
\eqn\SigmaGR{ \sigma_{\rm abs}(\omega) = {\pi^2 \over 2} 
R^2 r_K^3 \omega^2
    {{\omega \over 2 T_L} \over 1 - \e^{-{\omega \over 2 T_L}}}
\left (1 + {\omega^2 \over 16 \pi^2 T_L^2} \right )
\ .}
Using  that in the extremal limit 
$$ 
T_L = {r_K \over \pi r_1 r_5 } \ , 
$$ 
and remembering that we were only able to do the classical calculation
for $r_1=r_5=R$, we see that the D-branes and GR match if we take the
effective string tension to be
  \eqn\tension{
   T_{\rm eff} = {1\over 2 \pi R^2 } = {1 \over 2 \pi \alpha' g n_5} \ ,
  }
 where we have restored the dependence on $\alpha'$ that we have been
suppressing since \UnitChoice.  This value for $T_{\rm eff}$ is
precisely equal to the tension of the `fractionated' D-string moving
inside $n_5$ 5-branes \refs{\ms,\juan}. This is a highly non-trivial
independent check on the applicability of the effective string model
to fixed scalars, and also on the idea of D-string `fractionation!'
 
Another interesting comparison to be made is for near-extremal black
holes.
For $\omega,T_R\ll T_L$ but with ratio of $\omega$ to $T_R$ otherwise
arbitrary, \SigmaD\ becomes, using \tension\ for the tension,
\eqn\neare{
 \sigma_{\rm abs} (\omega)=  {\pi^2\over 2} R^2 r_K^3
(\omega^2+ 4 \pi^2 T_H^2) \  . } 
This is in exact agreement with the absorption cross-section 
on non-extremal black holes \Interp\ computed using general
relativity.

For the fixed scalar the coupling to $(\del X)^2$ is absent from
the D-brane action, and the cross-section we found is the leading
effect. For an ordinary `decoupled' scalar, such as the 6-d dilaton,
both terms are present. So, the cross-section  computed   above should be
part of the correction to the leading effect  determined  in \mast.
This is an interesting topic for future investigation.


\newsec{Conclusions}

Let us try to recapitulate in a few words what it has taken many
equations for us to state.  The main thrust of the paper has been to
explore the behavior of the type of fixed scalar studied earlier in
\refs{\gibb,\gkk,\fkk}, and most recently in \kr\ -- but now in the
context of five-dimensional black holes that can be modeled by bound
states of D1-branes and D5-branes \refs{\sv-\mast}.  For the most part
we have focused on the fixed scalar $\nu$ which corresponds to the
volume of the internal four-torus as measured by the string metric.
Through an interesting interplay between semi-classical computations
(where the basic theory is well known but analytically intractable in
general) and D-brane computations (where the theory is less well known
but very tractable), we have arrived at a general formula \SigmaD\ for
the cross-section for low-energy fixed scalars to be absorbed into the
black hole.

The absorption cross-section \SigmaD\ has a much richer and more
interesting functional form than the simple $\omega^2$ dependence
found in \kr.  Even in the simple limit $\omega,T_R \ll T_L$ in which
comparison calculations between GR and D-branes were initially
performed \refs{\dmw,\dm}, the fixed scalar cross-section goes not as
$\omega^2$ but as $\omega^2 + \kappa^2$, where $\kappa = 2 \pi T_H$ is
the surface gravity at the horizon.  While we have derived the
expression \SigmaD\ in full generality only in the D-brane picture, we
have demonstrated that it agrees with semi-classical calculations of
the cross-section in the two regimes: one regime reproduces this novel
$\omega^2 + \kappa^2$ behavior, while the other deals with absorption
into extremal black holes.  Because the equations for the gravitational 
perturbations and fixed scalar perturbations 
 couple unless two of the three charges, e.g., the 1-brane charge 
 and  the
5-brane charge,  are equal to each other,
 our semi-classical computations are limited to the
equal charge case (similar equal-charge assumption 
was used  in  $D=4$  case  in \kr). 
 Modulo this limitation, we have
confidence that a full greybody factor computation along the lines of
\mast\ would reproduce the general result \SigmaD.

One of the reasons why the extension of  the
semi-classical calculations to unequal 1-brane and 5-brane charges
(but with both still  greater than the third charge, 
$P,Q \gg Q_K$,   to remain in the dilute gas region)
would be interesting,   is that the D-brane computations 
involve one free
parameter, the tension $T_{\rm eff}$ of the effective string, which
can be read off from a comparison with a semi-classical calculation.
The expectation, based on the work of \refs{\ms,\juan} and on the
arguments given at the beginning of section~3, is that $T_{\rm eff} =
{1 \ov 2 \pi \alpha' g n_5}$.  Our work confirms this relation when the
1-brane and 5-brane charges are equal. 
 What a semi-classical calculation with
unequal charges should confirm  is that 
 $T_{\rm eff}$ is independent of the number of 1-branes. 

Although our ultimate goal has been to demonstrate a new agreement
between semi-classical GR and a perturbative treatment of the
effective string, we have along the way studied interesting facets of
both formalisms separately.  On the D-brane side, we have been forced
to go beyond the leading quadratic terms in the expansion of the DBI
action and examine terms quartic in the derivatives of the string 
collective coordinate fields $X^i$.  As we argued in
section~3, the generic form of the quadratic terms is practically
inevitable given the invariances of the problem.  But the decoupling
of the fixed scalar from quadratic terms and the precise form of its
coupling to quartic terms is a signature of the DBI action.  The
agreement between the D-brane and GR cross-sections for fixed scalars
is thus a more stringent test of the DBI action than the agreements
obtained previously \refs{\dm,\us,\mast} for ordinary scalars.

{}From the open string theory point of view,  the $(\del X)^2$ term 
in the  D-string action \nam,\expa\ originates 
 upon dimensional
reduction from the 
$F^2_{\m\n}$  term in  the $D=10$ Born-Infeld  action, while 
the $(\del X)^4$ terms correspond to  the $F^4_{\m\n}$ -terms.
It is amusing to note that the  fixed 
scalars, which are  coupled  to  the 
 Maxwell terms of the closed string  vector fields  in the 
space-time  effective action \acti, thus 
do not couple to the Maxwell term of the open string vector field 
in the effective D-string  action,  while  
 exactly the opposite is true for the 
`decoupled' scalars.
It is thus the $F^4_{\m\n}$ -terms in the DBI action 
(which are important also in some other contexts) that 
are effectively responsible  for the leading contribution 
to the cross-section of fixed scalars.

At the  relevant  $(\del X)^4$ order,  
the  processes involving fermionic excitations of
the effective string contribute in the same way  as purely bosonic
processes.  Fortunately,  the  coupling of
bosonic excitations to the fixed scalar field 
  predicted by the DBI action is of a particularly
simple form, $T^X_{++} T^X_{--} \nu(x)$, which admits an obvious
generalization to include fermions: $T^{\rm tot}_{++} T^{\rm tot}_{--}
\nu(x)$.  Obtaining precise agreement with GR using this coupling and
the normalization of $T_{\rm eff}$ as in 
\refs{\ms,\juan} may be
viewed   as  determining a partial supersymmetrization of the 
effective string action
via D-brane spectroscopy.

On the GR side, we have to some extent systematized the study of
 spherical  black hole configurations, including  spherically symmetric
   perturbations
around the basic $D=5$ black hole with three charges,  
  by reducing the problem to an effective  two-dimensional one. 
 For time-independent configurations, this  gives a 
straightforward derivation of the basic black hole solution. 
 We were
disappointed to find, however,  that,
 despite  relative  simplicity of the
effective two-dimensional theory compared to the full supergravity
equations, it still leads  to complicated  coupled
differential equations for  time-dependent fluctuations
around the static  solution.  So far, we have been able to extract simple
equations from the intractable general case only when some pair of
charges are equal.  Then the background value of one fixed scalar
becomes constant and its fluctuations decouple from the other fields,
leading to a non-extremal five-dimensional  generalization of the
equation studied in \kr.  Similar two-dimensional effective theory
techniques with similar equal charge limitations were  applied to
the basic four-dimensional black hole with four charges. 
 In this paper, we
have taken the four-dimensional calculations only far enough to see
that fixed scalars whose background values become constant when three
of the four charges are equal have an absorption cross-section with
the characteristic $\omega^2 + \kappa^2$ dependence.

One final comment is that we have focused almost exclusively on
absorption rather than Hawking emission.  This is not because Hawking
emission is any more difficult, but rather because agreement between
the semi-classical Hawking calculation and the D-brane result is
inevitable once a successful comparison of absorption cross-sections
is made.  To see this, one must only note that detailed balance
between emission and absorption is built into the Hawking calculation
and that it can be checked explicitly in the D-brane description.
Once detailed balance is established in both descriptions, it
obviously suffices to check that the absorption cross-section agrees
between the two in order to be sure that emission rates must agree as
well.


\bigbreak\bigskip\bigskip\centerline{{\bf Acknowledgements}}\nobreak

  We would like to thank S.~Mathur for useful discussions.  C.G.C.{}
and I.R.K.{} were supported in part by DOE grant DE-{}FG02-91ER40671. 
I.R.K.{} was also supported by the NSF Presidential Young Investigator
Award PHY-9157482 and  the James S.{} McDonnell {}Foundation grant
No.{} 91-48.  S.S.G.{} thanks the Hertz Foundation for its support.
A.A.T.{} would like to acknowledge the support of PPARC, the European
Commission TMR programme ERBFMRX-CT96-0045  and NATO grant CRG 940870.

\bigskip

\noindent 
 {\bf Note added:} After the completion of this paper, one of us
(I.R.K.) and M.~Krasnitz succeeded in deriving the general greybody
factor \SigmaDbrane\ from a classical general relativity absorption
calculation~\kk.  Complete agreement between general relativity
and the effective string model is achieved for the effective string
tension derived in \tension.

\vfill\eject


\appendix{A}{ }

The effective string action \expa\ may be used to classify
various scalar fields according to their coupling to
the black hole. While this action makes it clear that the
`fixed scalar' $\nu$ couples differently from the ordinary scalars,
$\bar h_{ij}$ and $\phi$, we also observe that there are scalars
with yet different properties, such as
$h_{5i}$. The purpose of this Appendix is to
calculate what their coupling to the effective string, given in
\iit, implies for the absorption rate.
We see that $h_{5i}$ couples either to two left-movers and
one right-mover or to two right-movers and one left-mover.
The absorption processes due to the first type of coupling are
$scalar \rightarrow L+L+R$ and $scalar+L \rightarrow L+R$.
The relevant matrix element between properly normalized states is 
found to be
  \eqn\MatrixEl{
   {\kappa_5\over \sqrt {T_{\rm eff}}} 
\sqrt{ q_1 p_1 p_2\over \omega}\  \ .
  }
Adding up the absorption rate for the two processes gives
\eqn\llr{ \eqalign{&
{3\kappa_5^2 L_{\rm eff} \over 8\pi T_{\rm eff} \omega} 
{\omega\over 1- \e^{-{\omega\over 2 T_R}} }
       \int_{-\infty}^\infty d p_1 d p_2 \,
        \delta\left( p_1 + p_2 -{\omega\over 2} \right)
        {p_1 \over 1 - \e^{-{p_1\over T_L}}}
        {p_2 \over 1 - \e^{-{p_2\over T_L}}}\cr
= &
{\kappa_5^2 L_{\rm eff} \over 128 \pi T_{\rm eff}}
        {\omega \over
         \left(1- \e^{-{\omega\over 2 T_L}} \right)
         \left(1- \e^{-{\omega\over 2 T_R}} \right)
        }
        \left( \omega^2 + 16 \pi^2 T_L^2 \right) \ . 
}}
The absorption rate due to the processes
$scalar \rightarrow R+R+L$ and $scalar+R \rightarrow R+L$
is calculated analogously, so that the total absorption rate for a
scalar $h_{5i}$ is
\eqn\totalrate{\Gamma_{\rm abs}(\omega)=
{\kappa_5^2 L_{\rm eff} \over 64 \pi T_{\rm eff}}
        {\omega \over
         \left(1- \e^{-{\omega\over 2 T_L}} \right)
         \left(1- \e^{-{\omega\over 2 T_R}} \right)
        }
        \left( \omega^2 + 8 \pi^2 T_L^2 + 8 \pi^2 T_R^2\right)\ . 
}
Now the classical absorption cross-section is found from the relation
\SigmaGamma. Using \klrel\ and \tension, and setting $r_1=r_5=R$,
we obtain
\eqn\offdiag{
\sigma_{\rm abs}(\omega) = { \pi^3  R^6 \over 8}
{\omega \left (\e^{\omega\over T_H} - 1 \right ) \over   
\left (\e^{\omega\over 2 T_L} - 1\right )
\left (\e^{\omega\over 2 T_R} - 1\right ) }
(\omega^2 + 8 \pi^2 T_L^2 + 8 \pi^2 T_R^2)
}
This cross-section has a number of interesting properties.
In the limit $\omega\rightarrow 0$ it approaches
\eqn\smallom{\sigma_{\rm abs}(0) =  \pi^2 (2 r_K^2 + r_0^2) \sqrt {r_K^2 + r_0^2} \ . 
}
This is clearly different from the behavior found for the fixed scalar
$\nu$: the $\omega=0$ cross-section of $h_{5i}$ does not vanish
at extremality. The expression
 \smallom\ is also different from the cross-section 
 $\sigma_{\rm abs}(0)=A_{\rm h}$ which is found for ordinary scalars. We conclude that
$h_{5i}$ is neither the fixed scalar of the type exhibited in \kr\
nor the ordinary massless scalar.

It is interesting to see what \offdiag\ reduces to in the extremal
limit where $r_0$ and $T_R$ are sent to zero. Here we find
\eqn\another{ \sigma_{\rm abs}(\omega)= 2\pi^2 r_K^3 
{ {\omega\over 2 T_L} \over 1 - \e^{-{\omega\over 2 T_L}} }
\left (1 + {\omega^2\over 8 \pi^2 T_L^2} \right )
\ . }
Using the parameter $\eta$ introduced in \EtaDef\ this
is equal to
\eqn\yet{\sigma_{\rm abs}(\omega)= 2\pi^2 r_K^3 
\left (1 + 2 \eta^2 \right )
        \df{2 \pi \eta}{\e^{2 \pi \eta} - 1}
  \left[ 1 + O(r_K^2/ R^2) \right] \ .
}
It would be very interesting to compare the cross-section
\totalrate\ calculated for $h_{5i}$ using the effective string methods
to the corresponding classical GR cross-section.
The calculation of the latter is a rather complicated exercise
 which we postpone for the future.

\vfill\eject

\listrefs
\end